\documentclass[12pt,a4paper]{article}
\usepackage[english]{babel}
\usepackage{graphicx,rotating,epsfig}
\newcommand{\GeV} {\, {\rm GeV}}
\newcommand{\TeV} {\, {\rm TeV}}
\begin{document} 

\begin{center} 
\boldmath
\textbf{\large MEASUREMENT OF TRIPLE GAUGE BOSON COUPLINGS
    AT AN $e\gamma$ - COLLIDER}
\unboldmath
\end{center}
\begin{center} 
{\textbf{K.M\"{o}nig, J.Sekaric}} 
\end{center}
\begin{center} 
{Deutsches Elektronen-Synchrotron DESY \\
Platanenallee 6, 15738 Zeuthen, Germany}
\end{center}
%%%%%%%%%%%%%%%%
\begin{abstract}
%%%%%%%%%%%%%%%%
  If no light Higgs boson exists, the interaction among the gauge bosons
  becomes strong at high energies ($\sim 1 \, {\rm TeV}$). The effects of
  strong electroweak symmetry breaking (SEWSB) could manifest themselves
  indirectly as anomalous couplings before they give rise to new physical
  states like resonances. Here a study of the measurement of trilinear gauge
  couplings is presented looking at the hadronic decay channel of the W boson
  at an $e\gamma$ - collider. A sensitivity in the range of $10^{-3}$ to
  $10^{-4}$ can be reached depending on the coupling under consideration.
\end{abstract}
%%%%%%%%%%%%%%%%%%%%%%%%%%%%%%%%%%%%%%%%%%%%%%%%%%%%%%%%%%%%%%%%%%%%%%%
\section{Introduction}
%%%%%%%%%%%%%%%%%%%%%%%%%%%%%%%%%%%%%%%%%%%%%%%%%%%%%%%%%%%%%%%%%%%%%%%
The measurement of trilinear gauge couplings (TGCs) at a photon collider (PC)
\cite{tdr6} gives the possibility to study the bosonic sector of the Standard
Model (SM). Due to the non-Abelian nature of the gauge group which describes
the electroweak interactions, it is predicted that the gauge bosons interact
among themselves, giving rise to vertices with three or four gauge bosons.
Each vertex is described by dimensionless couplings, denoted as TGCs or QGCs
(triple or quartic gauge couplings) with a strength obtained in the SM
applying the ${SU(2)_{L}}\times{U(1)_{Y}}$ gauge symmetry. Possible deviations
from the values predicted by the SM, that could occur at high energies ($\sim
1 \, {\rm TeV}$), may indicate a signal of New Physics (NP) beyond the SM. In
this case the SM can be considered as a lower energy approximation of another
larger theory. The effects of this larger theory are contained in a
Lagrangian\footnote{ Chiral Lagrangian constructed in a similar way as the low
  energy QCD Lagrangian.}  expanded in power of $\frac{1}{\Lambda_{NP}}$,
where $\Lambda_{NP}$ is the scale of the NP:
\[
{\cal L}_{eff}=
\sum_{n \ge 0}\sum_{i}\frac{\alpha_{i}^{n}}{{\Lambda_{NP}}^{n}}{O}_{i}^{(n+4)}
\]
where the coefficients ${\alpha}_{i}$ are obtained from the parameters of the
high energy theory and parametrise all possible effects at low energy. 
The low-energy effective Lagrangian without a Higgs violates unitarity at a
scale %\footnote{
%  This estimate of ${\Lambda_{NP}}$ follows directly from analogy
%  with low energy QCD and chiral perturbation theory.} 
of 4$\pi\upsilon
\approx3\TeV$, so that new physics  should appear below this scale.
\par
Conventionally the trilinear gauge boson vertices, involving only W and
$\gamma$ bosons, are parametrised by the most general effective Lagrangian as
\cite{eff}:
\begin{eqnarray*}
{\cal L}_{TGC}^{WW} & = &
  -i e \left[
  g_{1}^{\gamma}V^{\mu}(W_{\mu\nu}^{-}W^{+\nu}-W_{\mu\nu}^{+}W^{-\nu})+
  {\kappa}_{\gamma}W_{\mu}^{+}W_{\nu}^{-}V^{\mu\nu}\right. \\
  & &
  +\frac{{\lambda}_{\gamma}}{M_{W}^{2}}V^{\mu\nu}W_{\nu}^{+\rho}W_{\mu\rho}^{-}
  \\
  & & 
  +i{g_{5}^{\gamma}}{\varepsilon_{\mu\nu\rho\sigma}}
  [({\partial}^{\rho}W^{-\mu})W^{+\nu}-{W^{-\mu}}
  ({\partial}^{\rho}W^{+\nu})]V^{\sigma}\\
  & &
  +ig_{4}^{\gamma}W_{\mu}^{-}W_{\nu}^{+}
  ({\partial}^{\mu}V^{\nu}-{\partial}^{\nu}V^{\mu})\\
  & & \left.
  -\frac{\tilde{{\kappa}_{\gamma}}}{2}W_{\mu}^{-}W_{\nu}^{+}
  {\varepsilon}^{\mu\nu\rho\sigma}V^{\rho\sigma}
  -\frac{\tilde{{\lambda}_{\gamma}}}{2M_{W}^{2}}
  W_{\rho\mu}^{-}W_{\nu}^{+}{\varepsilon}^{\nu\rho\alpha\beta}V_{\alpha\beta}]
  \right], 
%\label{eqn:s} \\
\end{eqnarray*}
where $M_{W}$ is the nominal $W^{\pm}$ mass, $V$ is the photon field, $W^\pm$
are the W fields, and the field tensors are given as
$W_{\mu\nu}={\partial}^{\mu}W^{\nu}-{\partial}^{\nu}W^{\mu}$ and
$V_{\mu\nu}={\partial}^{\mu}V^{\nu}-{\partial}^{\nu}V^{\mu}$.
$\varepsilon^{\alpha\beta\gamma\delta}$ is the fully antisymmetric
$\varepsilon$-tensor.  The seven coupling parameters of $\gamma$WW vertices
are grouped according to their symmetries as C and P conserving couplings
($g_{1}^{\gamma},{\kappa}_{\gamma}$ and ${\lambda}_{\gamma}$), C,P violating
but CP conserving couplings ($g_{5}^{\gamma}$) and CP violating couplings
($g_{4}^{\gamma},{\tilde{{\kappa}_{\gamma}}}$ and
${\tilde{{\lambda}_{\gamma}}}$). In the SM all couplings are zero except
$g_{1}^{\gamma}=1$ and ${\kappa}_{\gamma}=1$. As it was already mentioned,
deviations from the SM prediction, denoted as ${\Delta}{\kappa}_{\gamma}
(={\kappa}_{\gamma}-1)$ and ${\lambda}_{\gamma}$,
arise as a consequence of a new physics effect. Introducing
deviations of coupling parameters (``anomalous couplings'') from those given
in the SM, the previous Lagrangian in general describes non-renormalisable 
and unitarity violating interactions. % since it is not invariant under the
%${SU(2)_{L}}\times{U(1)_{Y}}$ gauge transformation.  
This analysis studies
the measurement of the C and P conserving couplings, ${\kappa}_{\gamma}$ and
${\lambda}_{\gamma}$, while the value of $g_{1}^{\gamma}$ is fixed by
electro-magnetic gauge invariance ($g_{1}^{\gamma}=1$). The other couplings
are assumed to vanish.
\par
The low energy effective Lagrangian for triple gauge boson vertices, in
non-linear realisation of the symmetry can be expressed in terms of the two
operators, ${\cal L}_{9L}$ and ${\cal L}_{9R}$ \cite{l91l91}, where
\begin{eqnarray*}
{\cal L}_{9L} & = &
ig_W\frac{L_{9L}}{{16\pi}^2}Tr[W^{\mu\nu}D_\mu\Sigma D_\nu \Sigma^{+}],\\
{\cal L}_{9R} & = &
ig_W^{'}\frac{L_{9R}}{{16\pi}^2}Tr[B^{\mu\nu}D_\mu \Sigma D_\nu \Sigma^{+}].
\end{eqnarray*}
${L}_{9L}$ and ${L}_{9R}$ are parameters expected to be of ${\cal O}(1)$
while $D_{\mu}\Sigma$ represents the ${SU(2)}\times{U(1)}$ covariant
derivative and $\Sigma=\exp(i{\vec{\omega}} \cdot {\vec{\sigma}}/{\upsilon})$
describes the Goldstone bosons with the built-in custodial $SU(2)_{C}$
symmetry. Taking the physical fields instead the Goldstone bosons the
following relation can be obtained:
$$
{\kappa}_{\gamma}=1+\frac{e^{2}}{{\sin^{2}\theta_{W}}}
\frac{{1}}{{32\pi}^{2}}(L_{9L}+L_{9R}).$$
Taking the operators of higher
dimension, ${\lambda}_{\gamma}$ is expected to be:
\[
{\lambda_{\gamma}}=(\frac{e^{2}}{{\sin^{2}\theta_{W}}})L_{\lambda}
\frac{{M_{W}}^{2}}{{\Lambda_{NP}}^{2}}.
\]
\par
If one assumes that any deviation from the SM values is induced by scattering
of Goldstone bosons\footnote{
  Longitudinal component of the gauge bosons, ${W_{L}}^{\pm}, Z_{L}$.} 
at high energy scales associated with spontaneous
symmetry breaking, this effective description without a physical Higgs boson
could explain the mass generation via the mechanism of SEWSB.
%\newpage
%%%%%%%%%%%%%%%%%%%%%%%%%%%%%%%%%%%%%%%%%%%%%%%%%%%%%%%%%%%%%%%%%%%%%%%
\section{Observables Sensitive to the Triple Gauge Couplings}
\begin{figure}[p]
\begin{center}
\epsfxsize=2.5in
\epsfysize=2.5in
\epsfbox{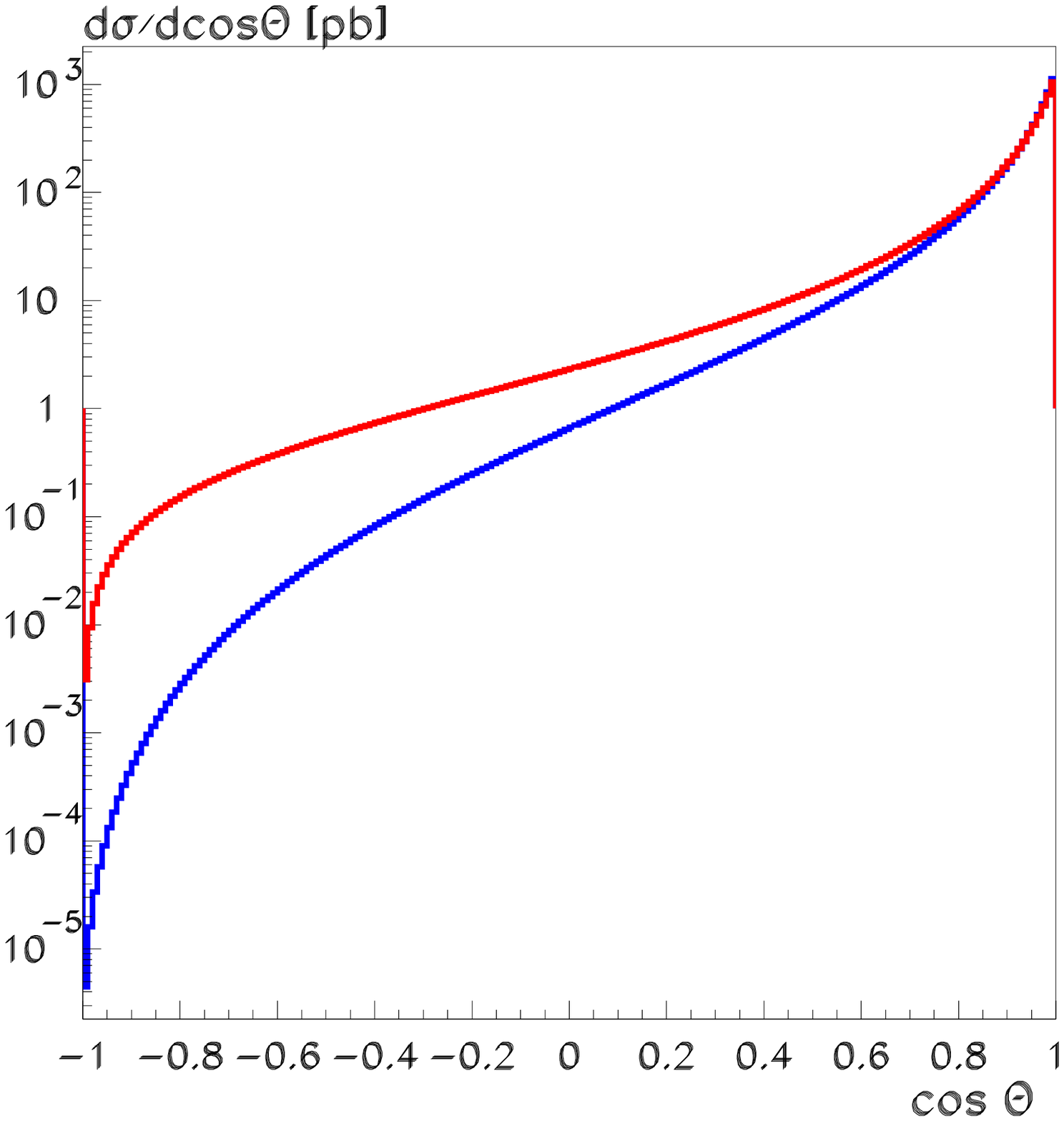}
\hspace{.25in}
\epsfxsize=2.in
\epsfysize=2.in
\epsfbox{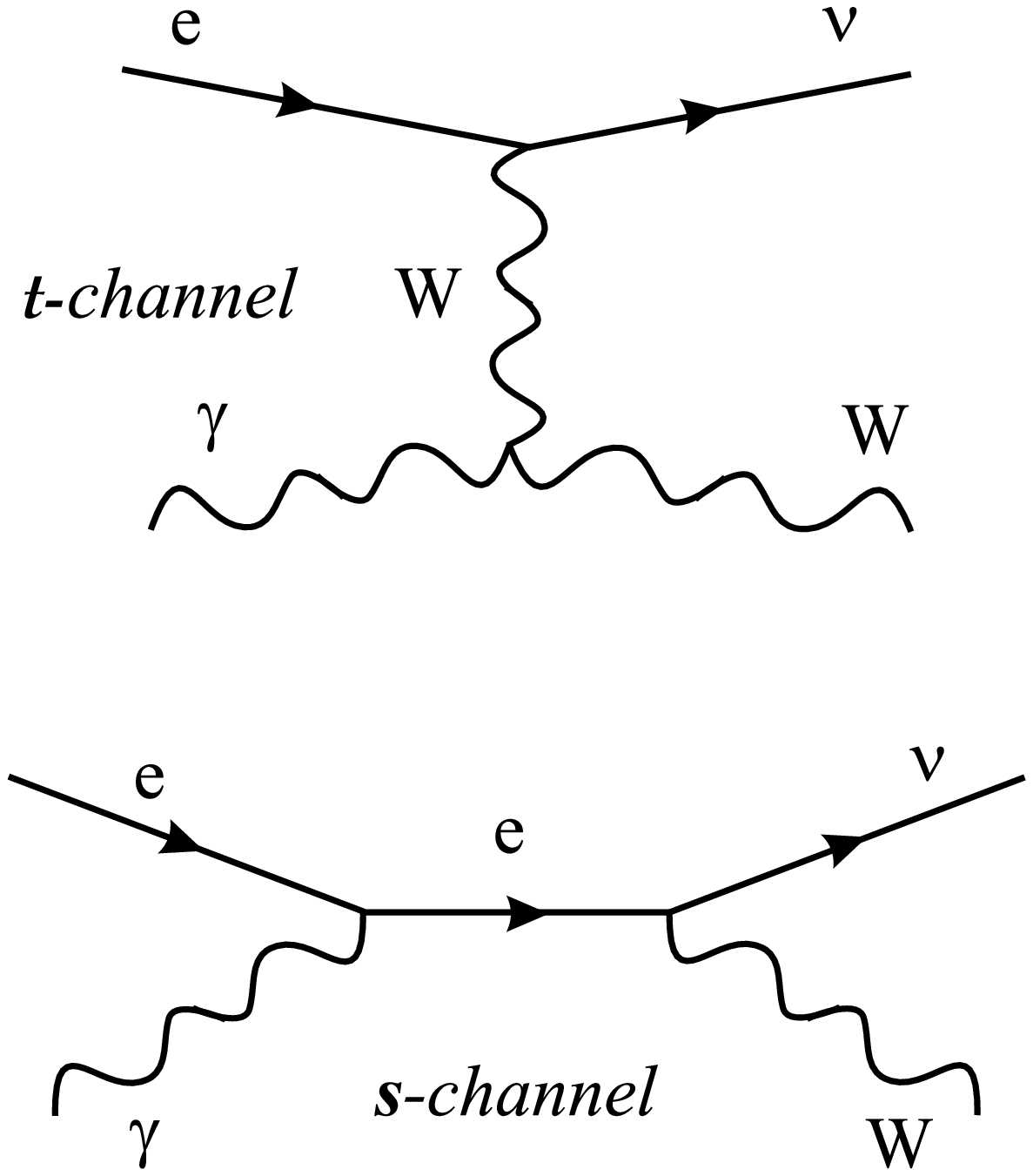}
\caption[bla]
{\textit{Left} (a): SM differential cross-section distributions for two
  different initial photon helicities - left-handed (\textit{red line-upper})
  and right-handed (\textit{blue line-lower}) 
  at $\sqrt{s_{e \gamma}} = 450 \GeV$. 
  The contribution from the
  \textit{s}-channel is visible for left-handed photons leading to a larger
  cross-section. \textit{Right} (b): Feynman diagrams contributing to $e^{-}
  \gamma \rightarrow W^{-} \nu_{e}$ with TGC contribution only through
  \textit{t}-channel \textit{W}-exchange.}
\label{fig:f1}
\end{center}
\end{figure}

\begin{figure}[p]
\begin{center}
\epsfxsize=2.5in
\epsfysize=2.5in
\epsfbox{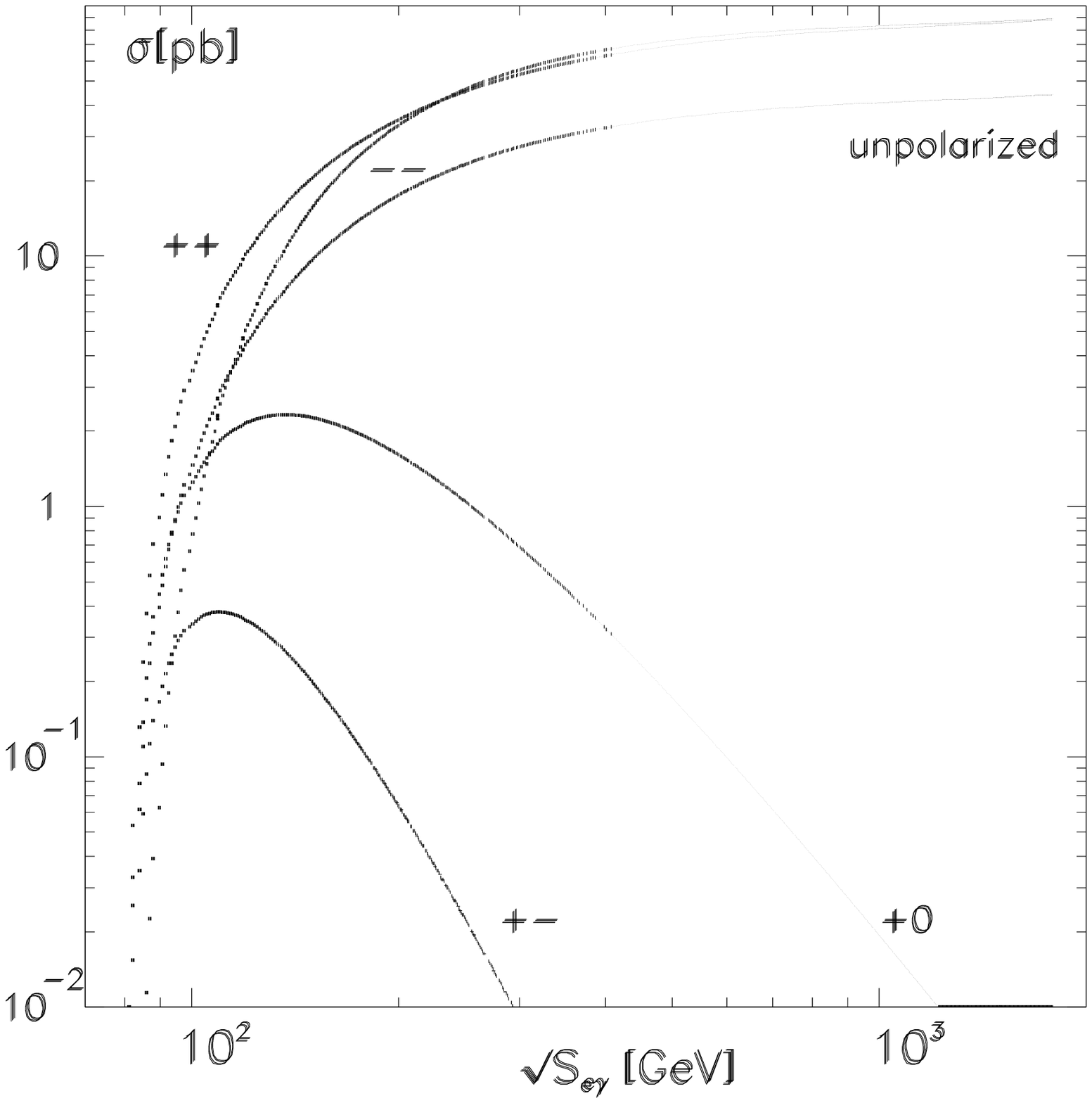}
\hspace{.25in}
\epsfxsize=2.5in
\epsfysize=2.5in
\epsfbox{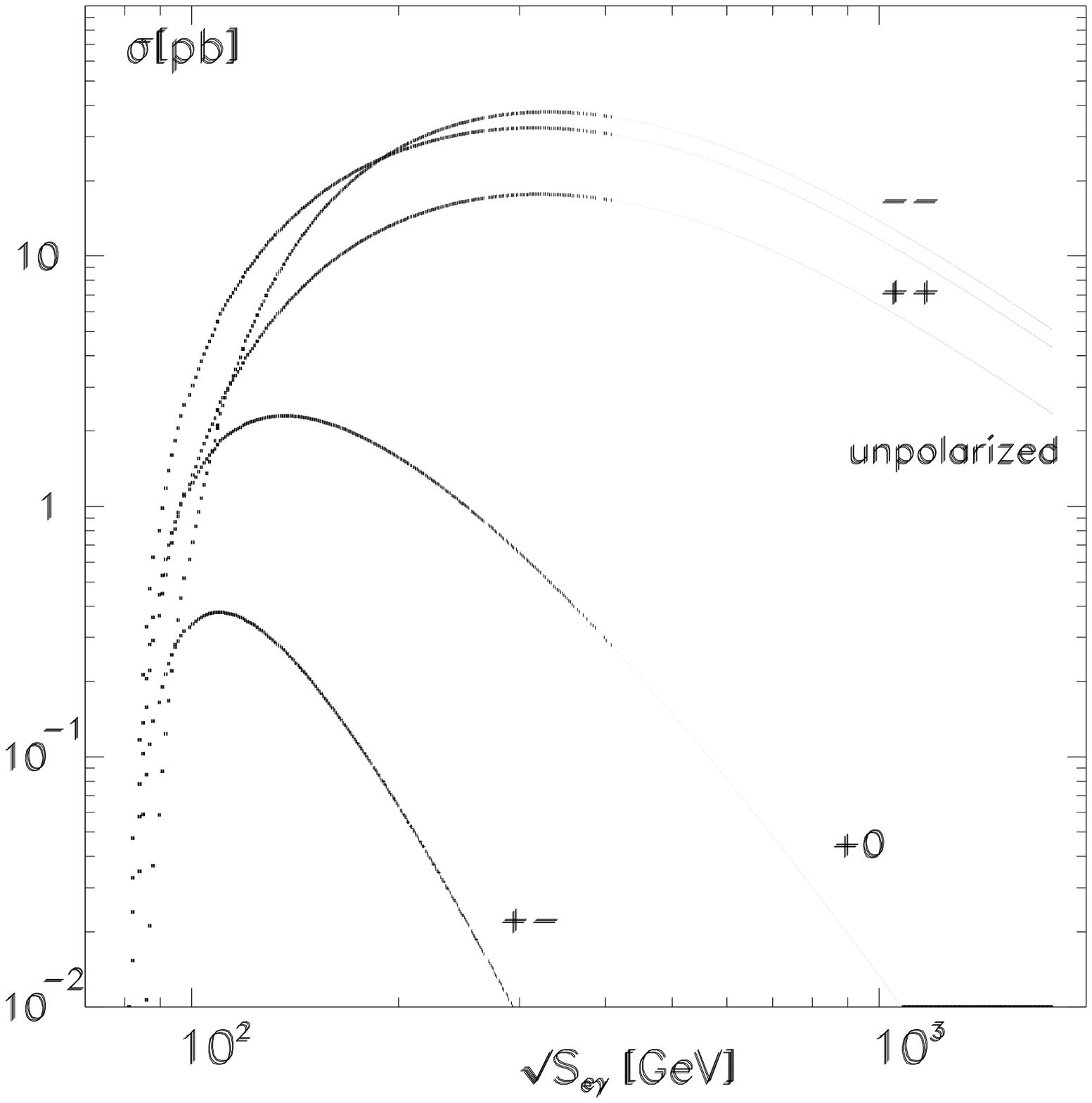}
\caption[bla]
{Total lowest-order cross-sections as a function of ${\sqrt{s_{e\gamma}}}$ for
  different boson polarisations assuming that the electron is left-handed.
  \textit{Left}: without an angular cut. \textit{Right}: with an angular cut
  $20^{\circ} \leq \theta \leq 160^{\circ}$. Notation: ($h_{\gamma},h_{W}$) =
  ($\gamma$ helicity, W helicity).}
\label{fig:f2}
\end{center}
\end{figure}

\begin{figure}[htb]
\begin{center}
\includegraphics[width=0.55\linewidth,bb=0 12 567 513]{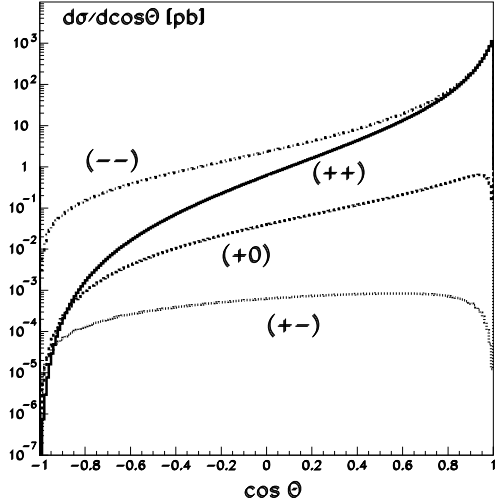}
%\epsfxsize=3.0in
%\epsfysize=3.0in
%\epsfbox{cross.eps}
\caption[bla]
{Contribution of each W helicity state for two different initial photon
  polarisations to the SM differential cross-section at
  $\sqrt{s_{e \gamma}} = 450 \GeV$. The angle
  $\theta$ is defined as the angle between the $\gamma$ beam and the outgoing
  W. Notation: ($h_{\gamma},h_{W}$) = ($\gamma$ helicity, W helicity).}
\label{fig:f3}
\end{center}
\end{figure}

%%%%%%%%%%%%%%%%%%%%%%%%%%%%%%%%%%%%%%%%%%%%%%%%%%%%%%%%%%%%%%%%%%%%%%%
We studied single W boson production in high energy $e\gamma$ collisions
($e^{-} \gamma \rightarrow W^{-} \nu_{e}$) and the sensitivity of some
observables like angular distributions, to the $\gamma$WW gauge boson
couplings. In $e\gamma$ collisions the TGCs contribute only through
\textit{t}-channel \textit{W}-exchange at the ${\gamma}$WW vertex 
as it is shown in Fig.~\ref{fig:f1}\,b. 
The beam electrons have to be left-handed since the W boson
does not couple to right-handed electrons. On the other hand, the photons
can be right-handed or left-handed. The differential cross-section for the two
different initial photon helicities is shown in Fig.~\ref{fig:f1}\,a. For
left-handed photons the \textit{s}-channel contribution leads to a higher
differential cross-section.  The contribution of each W helicity state to the
total cross-section for different centre-of-mass energies is shown in
Fig.~\ref{fig:f2}. The contribution of each W helicity state to the
differential cross-section is shown in Fig.~\ref{fig:f3}.

For the boson polarisations $(h_{\gamma},h_{W})=(-1,+1)$ and $(-1,0)$ the SM
amplitudes are equal to zero. Different W-helicity states are contained in the
differential cross-section distribution over the decay angle:
\[
{\frac{{d^2\sigma}}{d\cos\theta d\cos{\theta}_{1}}} =
{\frac{3}{4}}\left[{\frac{1}{2}}{\frac{{d{\sigma}_{T}}}
{d\cos\theta}}(1+{\cos^{2}}{\theta_{1}})+{\frac{{d{\sigma}_{L}}}
{d\cos\theta}}{\sin^{2}}{\theta_{1}}\right],
\]
where $\theta$ denotes the production angle of the W and $\theta_{1}$ denotes
the decay angle. ${\frac{{d{\sigma}_{T}}}{d{\cos \theta}}}$ is the
differential cross section for the production of transversally polarised
W-bosons distributed as ${(1+{\cos^{2}}{\theta_{1}})}$ and
${\frac{{d{\sigma}_{L}}}{d{\cos \theta}}}$ is the differential cross section
for longitudinal W production, distributed as ${\sin^{2}}{\theta_{1}}$.
\par
\begin{figure}[p]
\begin{center}
\epsfxsize=2.5in
\epsfysize=2.5in
\epsfbox{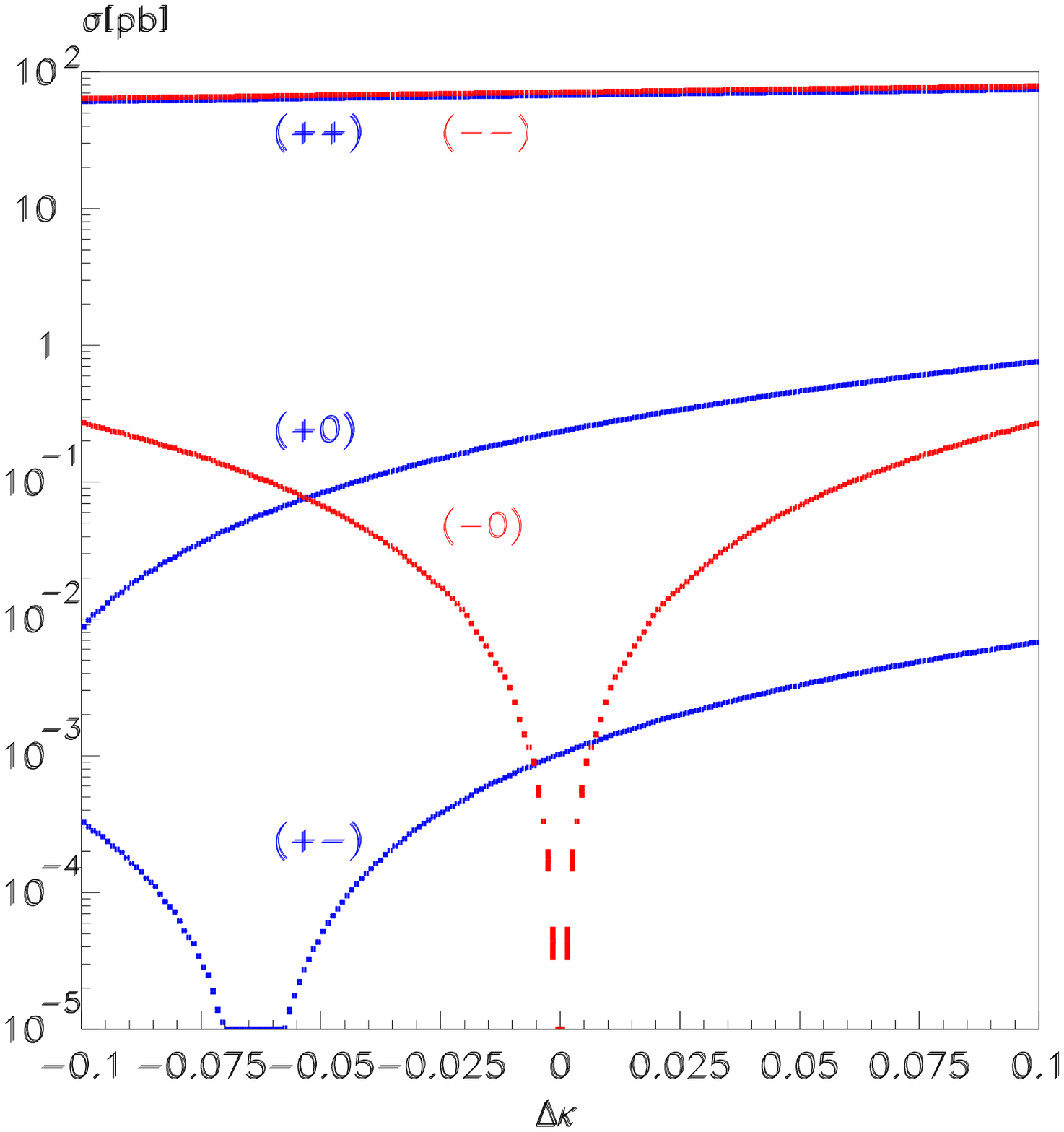}
\hspace{.25in}
\epsfxsize=2.5in
\epsfysize=2.5in
\epsfbox{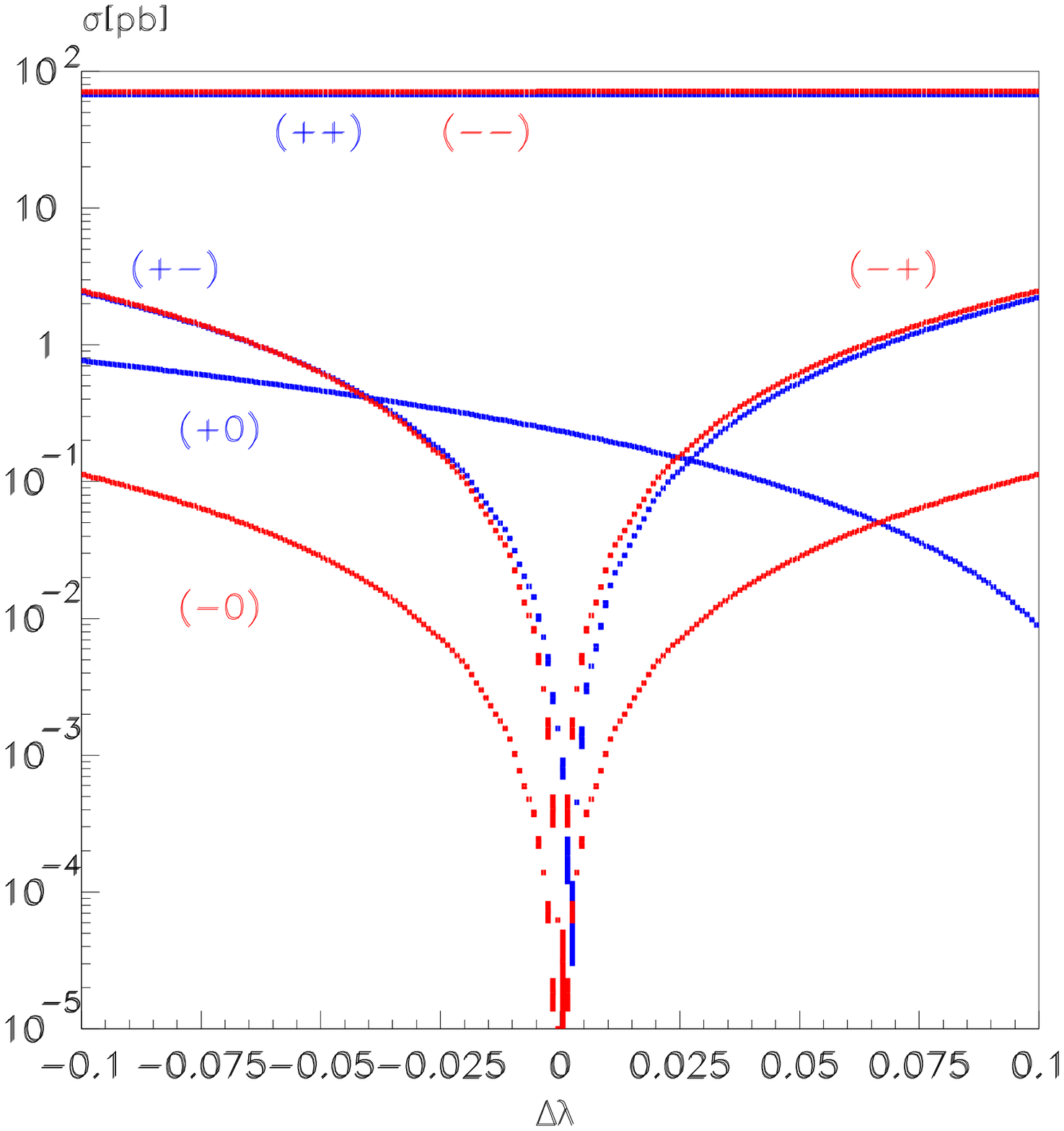}
\caption
{Contribution of different W helicity states for two different initial photon
  polarisations in the presence of anomalous couplings,
  ${\Delta}{\kappa}_{\gamma}$ (\textit{left plot}) and ${\lambda}_{\gamma}$
  (\textit{right plot}) at $\sqrt{s_{e \gamma}} = 450 \GeV$. 
  Notation: ($h_{\gamma},h_{W}$) = ($\gamma$ helicity, W helicity).}
\label{fig:f4}
\end{center}
\end{figure}
\begin{figure}[p]
\begin{center}
  \epsfxsize=2.5in \epsfysize=2.5in \epsfbox{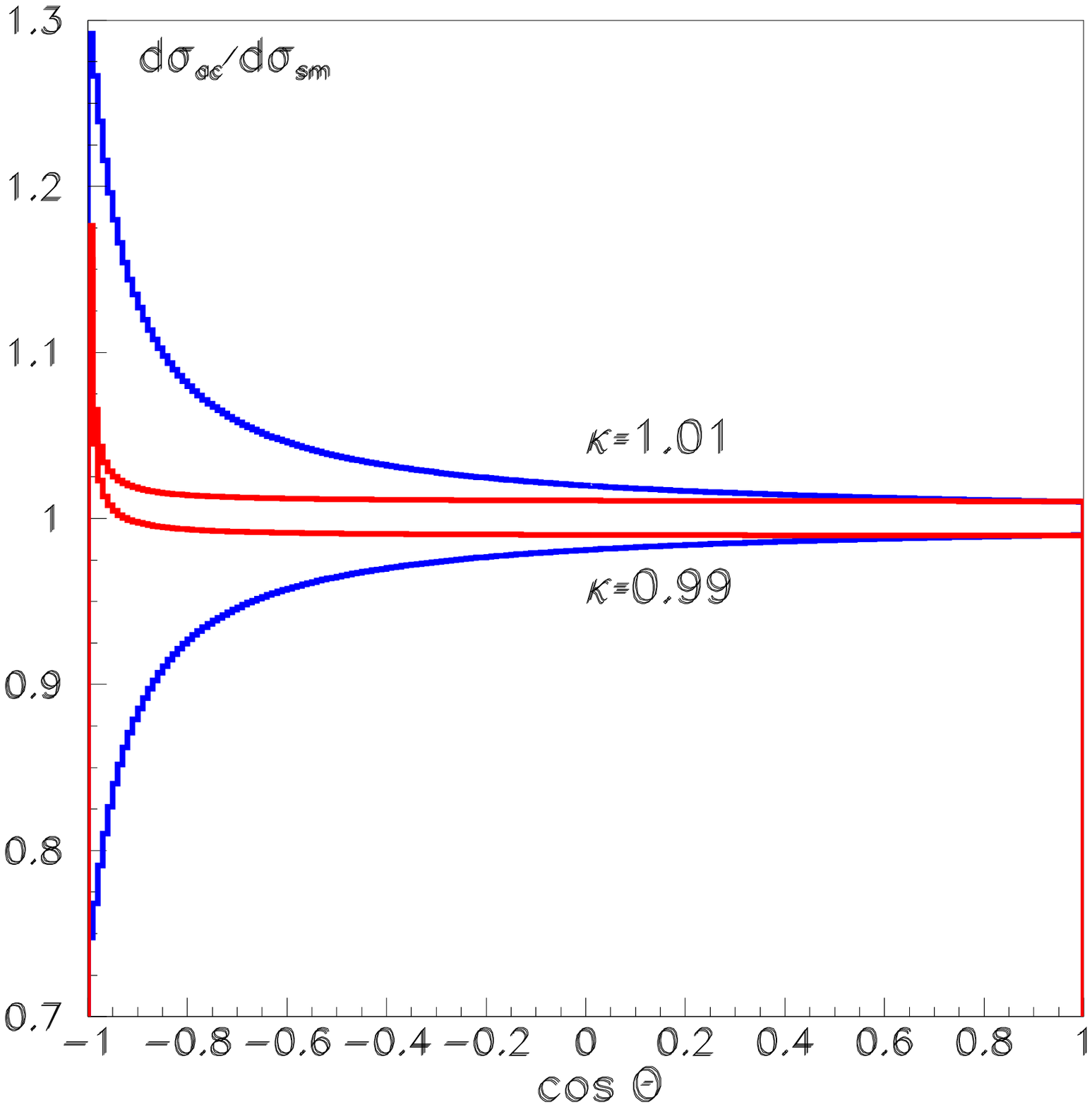} \hspace{.25in}
  \epsfxsize=2.5in \epsfysize=2.5in \epsfbox{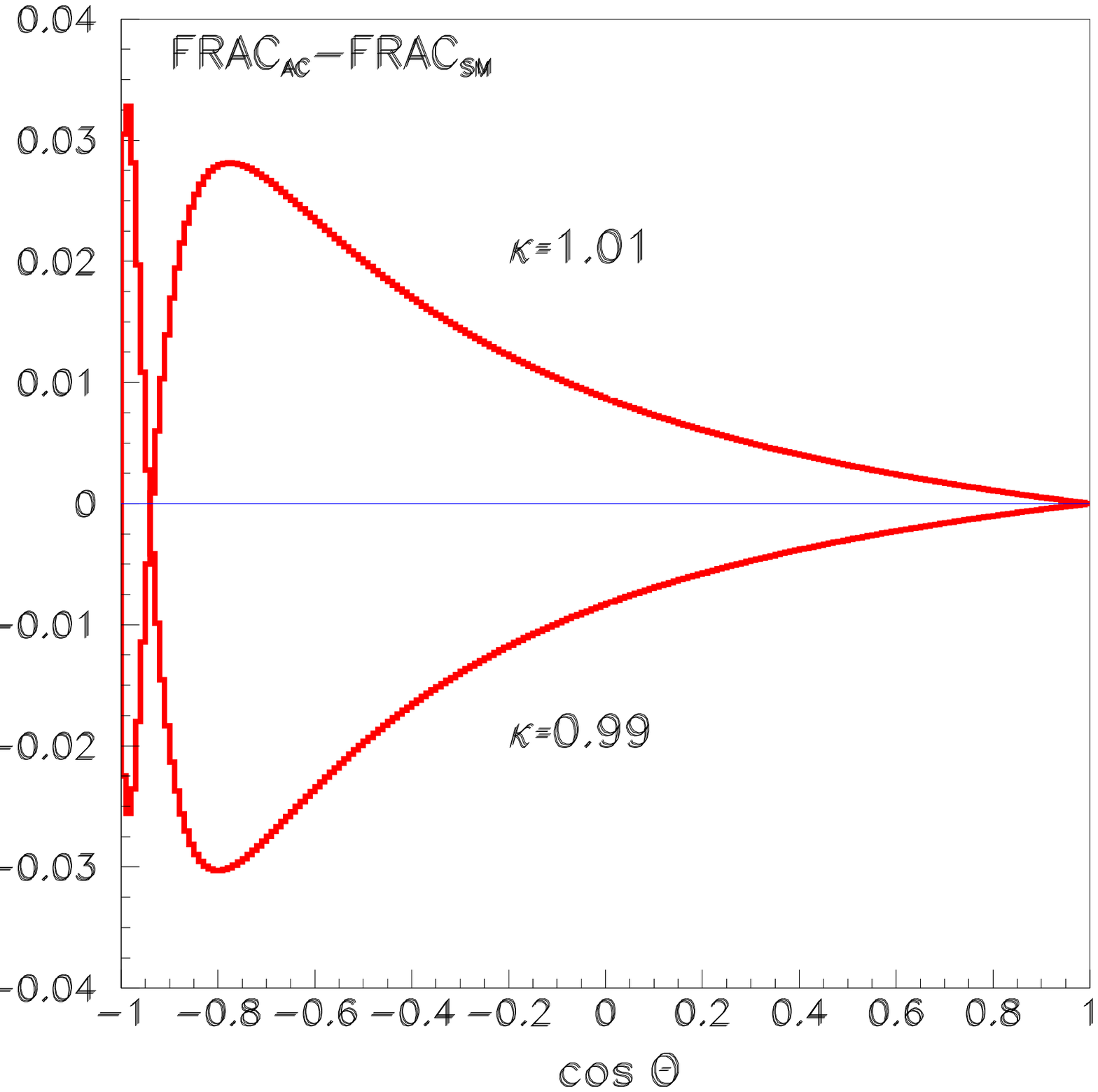}
\caption[bla]
{\textit{Left} (a): Differential cross-section in the presence of anomalous
  TGCs for both initial photon helicity states - left-handed
  (\textit{blue-outer lines}) and right-handed (\textit{red-inner lines}),
  normalised to their SM values at $\sqrt{s_{e \gamma}} = 450 \GeV$. 
  \textit{Right} (b): Deviation of longitudinal W fraction in presence of 
  anomalous TGCs from the SM for ${\Delta}{\kappa}_{\gamma}={\pm}0.01$
  at $\sqrt{s_{e \gamma}} = 450 \GeV$.}
\label{fig:f5}
\end{center}
\end{figure}

Anomalous TGCs affect both the total production cross-section and the shape of
the differential cross-section as a function of the W production angle. As a
consequence, distributions of W decay products are changed also. The relative
contribution of each helicity state of the W boson to the total cross-section
in the presence of anomalous couplings is shown in Fig.~\ref{fig:f4}.
Fig.~\ref{fig:f5}\,a shows that the differential cross-section distribution in
the backward\footnote{
  W production angle is defined as the angle between the
  photon and the W boson.} 
region is more sensitive to the presence of
anomalous TGC in the case of right-handed photons than for left-handed
ones. Fig.~\ref{fig:f5}\,b shows the $W_{L}$ fraction if there are
anomalous couplings and in the SM. The production of $W_{L}$ bosons in the
presence of anomalous couplings will differ from the SM. This behaviour comes
from the fact that the information about SEWSB can be obtained through the
study of Goldstone boson interactions which are the longitudinal component of
the gauge bosons. Differential cross-sections are calculated on the basis of
the formula given in \cite{dit} using helicity amplitudes in the presence of
anomalous couplings from \cite{yehu}.  In W production via $e\gamma$
collisions the favourable initial $\gamma$-$e$ helicity states are 
``right-left'' respectively. 
Because of the missing \textit{s}-channel electron-exchange in this state,
the W boson angular distributions show larger
sensitivity to TGCs in the backward region than in the case with initial
left-handed photons.
\newpage
%%%%%%%%%%%%%%%%%%%%%%%%%%%%%%%%%%%%%%%%%%%%%%%%%%%%%%%%%%%%%%%%%%%%%%%%%%%%%%
\section{Signal and Background Simulation}
%%%%%%%%%%%%%%%%%%%%%%%%%%%%%%%%%%%%%%%%%%%%%%%%%%%%%%%%%%%%%%%%%%%%%%%%%%%%%%
The energetic, highly polarised photons can be produced at a high rate in 
Compton backscattering of laser photons on high energy electrons
\cite{tdr6}. Setting opposite helicities for the laser photons and the beam
electrons the energy spectrum of the backscattered photons is
peaked at $\sim 80\,{\%}$ of the electron beam-energy. The
backscattered photons are highly polarised in this high energy region. 
With an integrated luminosity in the real mode of 
$71\, {\rm fb}^{-1}/{\rm year}$\footnote{
  A year is assumed to be $10^7\,{\rm s}$ at design luminosity.}
for
$\sqrt{s_{e \gamma}} > 0.8 \sqrt{s_{e \gamma}({\rm max})}$ \cite{tdr6},
$3 \cdot 10^{6}$ Ws per year can be produced in 
$e^{-} \gamma \rightarrow W^{-} \nu_{e}$
with hadronically decaying Ws, assuming $100\,{\%}$ detector acceptance.
In the parasitic mode the luminosity is even slightly higher.
\par
A photon collider can operate as a $\gamma\gamma$- or as 
an $e\gamma$-collider. 
$e\gamma$-collisions can be studied in two different modes - the real
and the parasitic one. In the real mode electrons from only one
electron beam are converted into high energy photons 
($e\gamma$-collider). If the electrons from both electron beams are converted
into high energy photons the $\gamma\gamma$-collider is realised and the
interactions between backscattered photons and unconverted electrons from both
sides can be used in the parasitic $e\gamma$ mode.
\par
\begin{figure}[htb]
\begin{center}
\epsfxsize=2.5in
\epsfysize=2.5in
\epsfbox{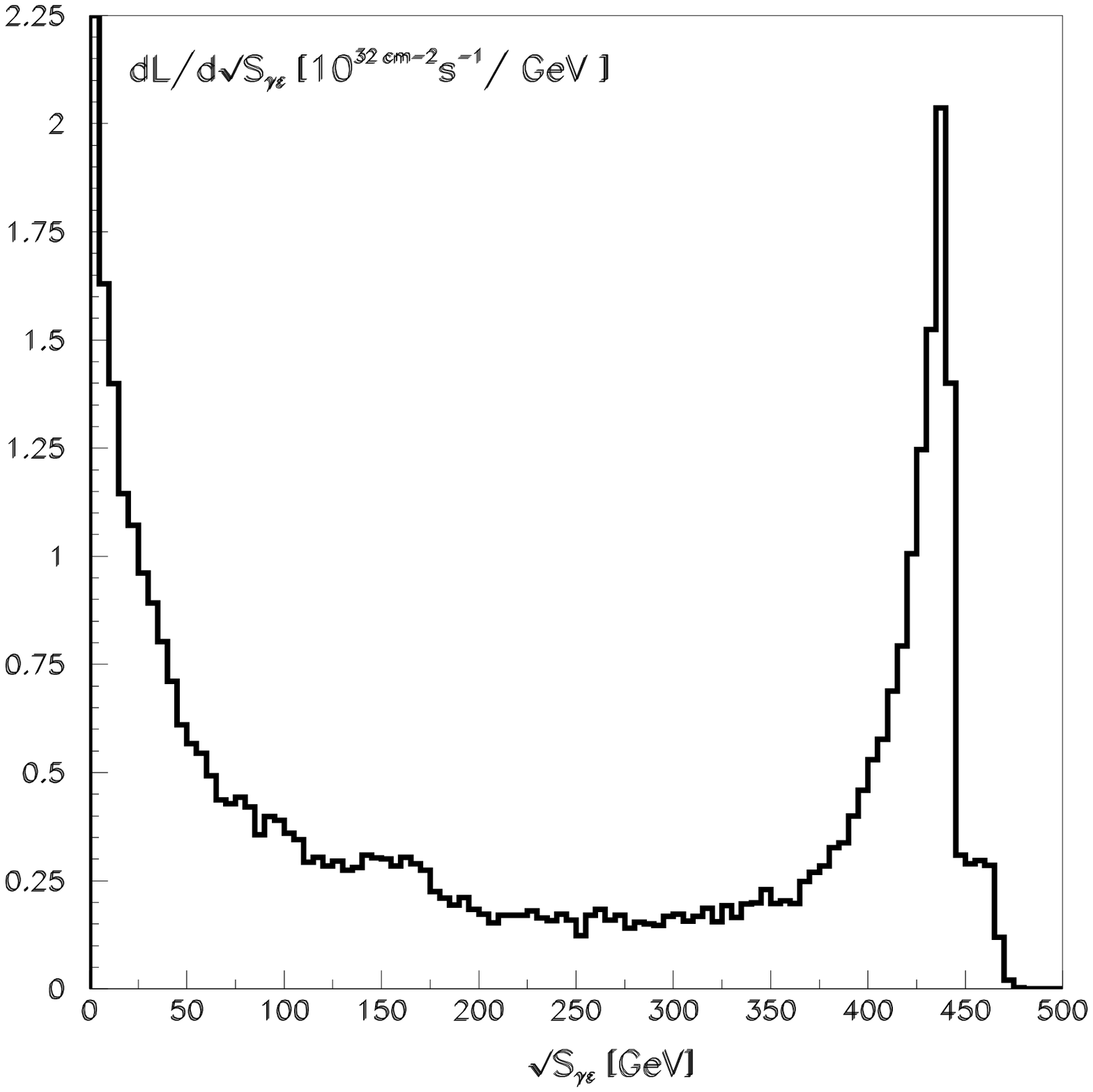}
\hspace{.25in}
\epsfxsize=2.5in
\epsfysize=2.5in
\epsfbox{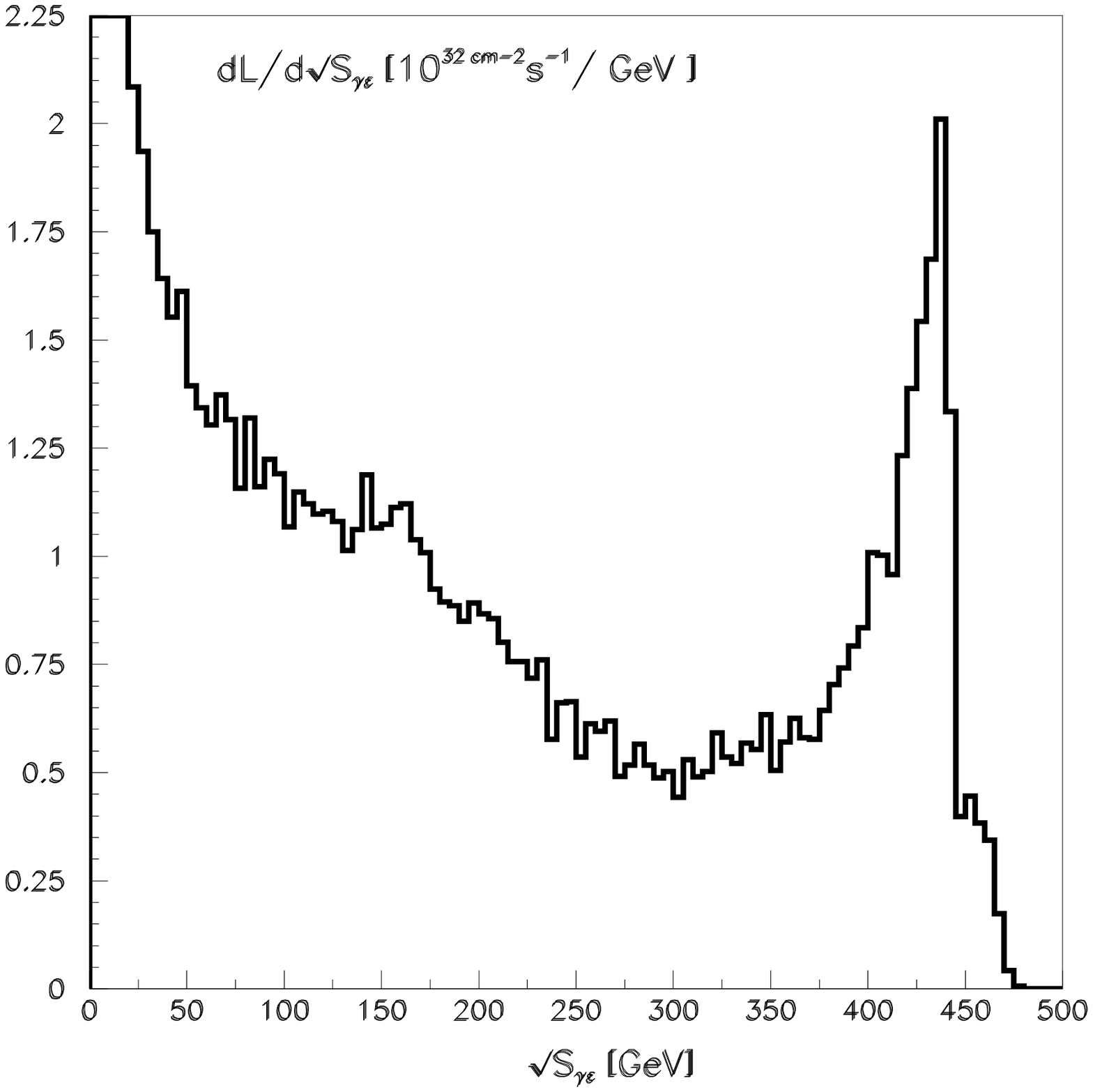}
\caption[bla]{$e\gamma$ luminosity spectra for the real (left) and parasitic
  (right) mode simulated with CIRCE2 for
  ${\sqrt{s_{ee}}=500}\,{\rm GeV}$.}
\label{fig:f6}
\end{center}
\end{figure}

The beam spectra for the different collider modes at
${\sqrt{s_{ee}}=500}\,{\rm GeV}$
are simulated using CIRCE2 \cite{circe2}.
CIRCE2 is a fast parameterisation of the spectra described in \cite{tdr6}
including multiple interactions and non-linearity effects.
The used spectra for the two modes are shown in Fig.~\ref{fig:f6}.

The response of the detector has been simulated with SIMDET V4 \cite{simdet4},
a parametric Monte Carlo for the TESLA $e^{+}e^{-}$ detector. It includes a
tracking and calorimeter simulation and a reconstruction of
energy-flow-objects (EFO)\footnote{
  Electrons, photons, muons, charged and
  neutral hadrons and unresolved clusters that deposit energy in the
  calorimeters.}. 
Only the EFOs with a polar angle above $7^{\circ}$ are taken
for the W boson reconstruction simulating the acceptance of the PC detector as
the only difference to the $e^{+}e^{-}$ detector \cite{ggparis}. 
W bosons are reconstructed using
the hadronic decay channel (BR=0.68). The signal and background events
are studied on a sample of events generated with WHIZARD \cite{whizard}.

The hadronic cross-section for ${\gamma\gamma}{\rightarrow}{\rm hadrons}$
events, within the energy range above $2\, {\rm GeV}$, is several
hundred nb \cite{pdg} so that ${\cal O}(1)$ events of this type are produced
per bunch crossing (pileup). These events are overlayed to the signal events.
Depending on the photon spectra the hadronic cross-section and the number of
hadronic events can be calculated using different models including real and
virtual photons \cite{thesis}. Since these events are induced by
\textit{t}-channel \textit{q}-exchange most of the resulting final state
particles  are distributed at
low angles.  

The informations about the neutral particles (\textit{neutrals})
from the calorimeter and charged tracks (\textit{tracks}) from the tracking
detector are used to reconstruct the signal and background events.  The
considered backgrounds depend on the two different modes of the
$e \gamma$-collider and for both modes result in a ${q}{\bar{q}}$-pair in the
final state.  Due to the different $\gamma\gamma$ luminosities in the two
$e\gamma$ modes, the pileup contribution to each mode is different - 1.2
events/BX for the real mode and 1.8 events/BX for the parasitic mode
\cite{schulte}. A large cross-section for W boson production
(${\sigma}_{pol}\sim45\,{\rm pb}$ for the \textrm{hadronic channel}) provides
an efficient separation of signal from background applying several successive
cuts.
\par
For the real mode the considered backgrounds are following:
\begin{enumerate}
\item ${e}{\gamma}{\rightarrow}{e}{Z} {\rightarrow}e {q}{\bar{q}}$,
  where the events are simulated with a kinematic cut which allows only
  production of electrons at low angles (below $15^\circ$). The preselection
  cut used to reduce the background contributing to this channel was to reject
  events with a high energetic electron ($\geq 100\, {\rm GeV}$) in the
  detector. By this cut $33\%$ of the background events are rejected not
  affecting the signal efficiency.
\item ${\gamma}(e^-){\gamma}{\rightarrow}{q}{\bar{q}}$,
  simulating the interaction between a real, high energy photon and a virtual
  bremsstrahlung photon.
\end{enumerate}
\par
Additional backgrounds considered for the parasitic mode are the following:
\begin{enumerate}
\item ${\gamma}{\gamma}{\rightarrow}{W}{W}$,
  where one W decays leptonically and the other W decays
  hadronically. To reduce the background contributing from
  this channel in each event we searched for a lepton in the detector with an
  energy higher than $5\,{\rm GeV}$. For these leptons a cone of $30^{\circ}$
  is defined around their flight directions and the energies of all particles
  (excluding the lepton) are summed inside the cone. Events with energies
  smaller than $20\,{\rm GeV}$ were rejected. This cut rejects $\sim70\%$ of
  the semileptonic WW background events, not affecting the signal efficiency.
\item ${\gamma}{\gamma}{\rightarrow}{q}{\bar{q}}$,
simulating the interaction between two real photons.
\end{enumerate}
%%%%%%%%%%%%%%%%%%%%%%%%%%%%%%%%%%%%%%%%%%%%
\subsection{Energy flow and Event Selection}
\begin{figure}[htb]
\begin{center}
\epsfxsize=2.5in
\epsfysize=2.5in
\epsfbox{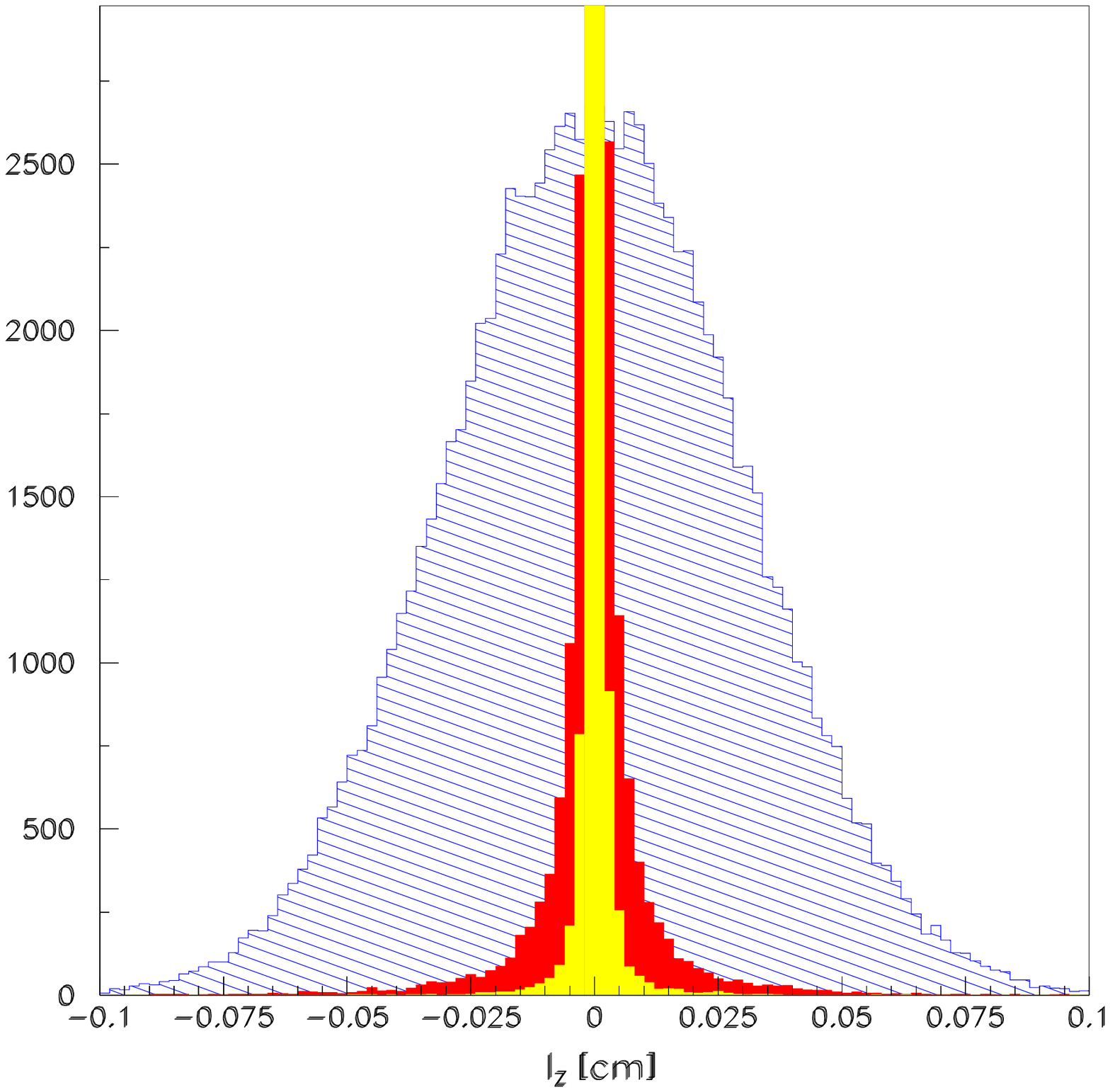}
\hspace{.25in}
\epsfxsize=2.5in
\epsfysize=2.5in
\epsfbox{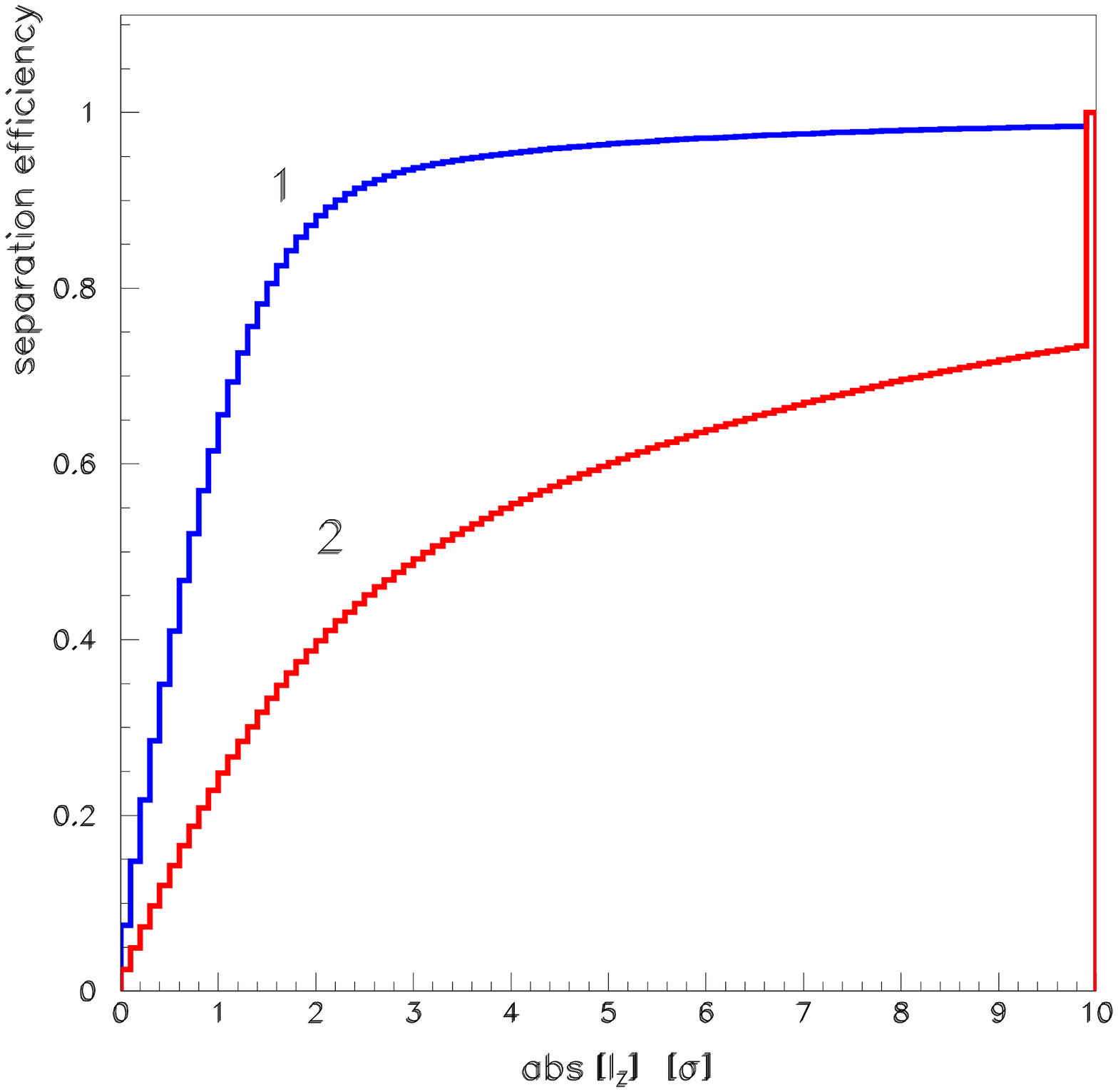}
\caption[bla]
{\textit{Left} (a): Primary vertex distribution along the beam axis
  (\textit{blue - hatched}) compared to the deviation of the reconstructed
  primary vertices for signal without (\textit{yellow-light}) and with pileup
  (\textit{red-dark}) tracks. \textit{Right} (b): Separation efficiency
  for signal (\textit{blue-1}) and pileup (\textit{red-2}) tracks for
  $\textit{$I_{X,Y}$}\leq{2\sigma}$.}
\label{fig:f7}
\end{center}
\end{figure}

%%%%%%%%%%%%%%%%%%%%%%%%%%%%%%%%%%%%%%%%%%%%
In order to minimise the pileup contribution to the high energy
signal tracks the first step in the separation procedure was to
reject pileup tracks as much as possible. The measurement of the impact
parameter of a particle along the beam axis with respect to the primary vertex
is used for this purpose. The beamspot length of 300$\mu{\rm m}$ for TESLA is
simulated and shown in Fig.~\ref{fig:f7}\,a, representing the primary vertex 
distribution of events along the \textit{z}-axis.

Using the precise measurements from the vertex detector first the primary
vertex of an event is reconstructed as the momentum weighted average z-impact
parameter\footnote{
  The z-impact parameter is defined as the z coordinate of the impact
  point in the $x-y$ plane.}, 
$I_Z$ of all tracks in the event. 
All impact parameters are then recalculated using
this primary vertex.
The reconstructed primary vertex distribution for signal with and without 
pileup tracks is also shown in Fig.~\ref{fig:f7}\,a. 
It can be seen that the distribution with pileup tracks is much broader 
than if there are only signal tracks.
The separation efficiency for a cut on
$|I_{Z}/\sigma|$ is shown in Fig.~\ref{fig:f7}\,b for tracks with a
transversal impact parameter $\textit{$I_{X,Y}$}$ less than ${2\sigma}$. If
one selects the tracks with \textit{$I_{Z}$} less than ${2\sigma}$, about
$\sim 60\,{\%}$ of the pileup tracks and only $\sim 5-10\,{\%}$ of the 
signal tracks are
rejected. All tracks with $\textit{$I_{X,Y}$}\geq{2\sigma}$ are accepted since
they could originate from a secondary vertex of a good track.
\par
\begin{figure}[htb]
\begin{center}
\includegraphics[width=0.48\linewidth,bb=0 5 567 542,clip]{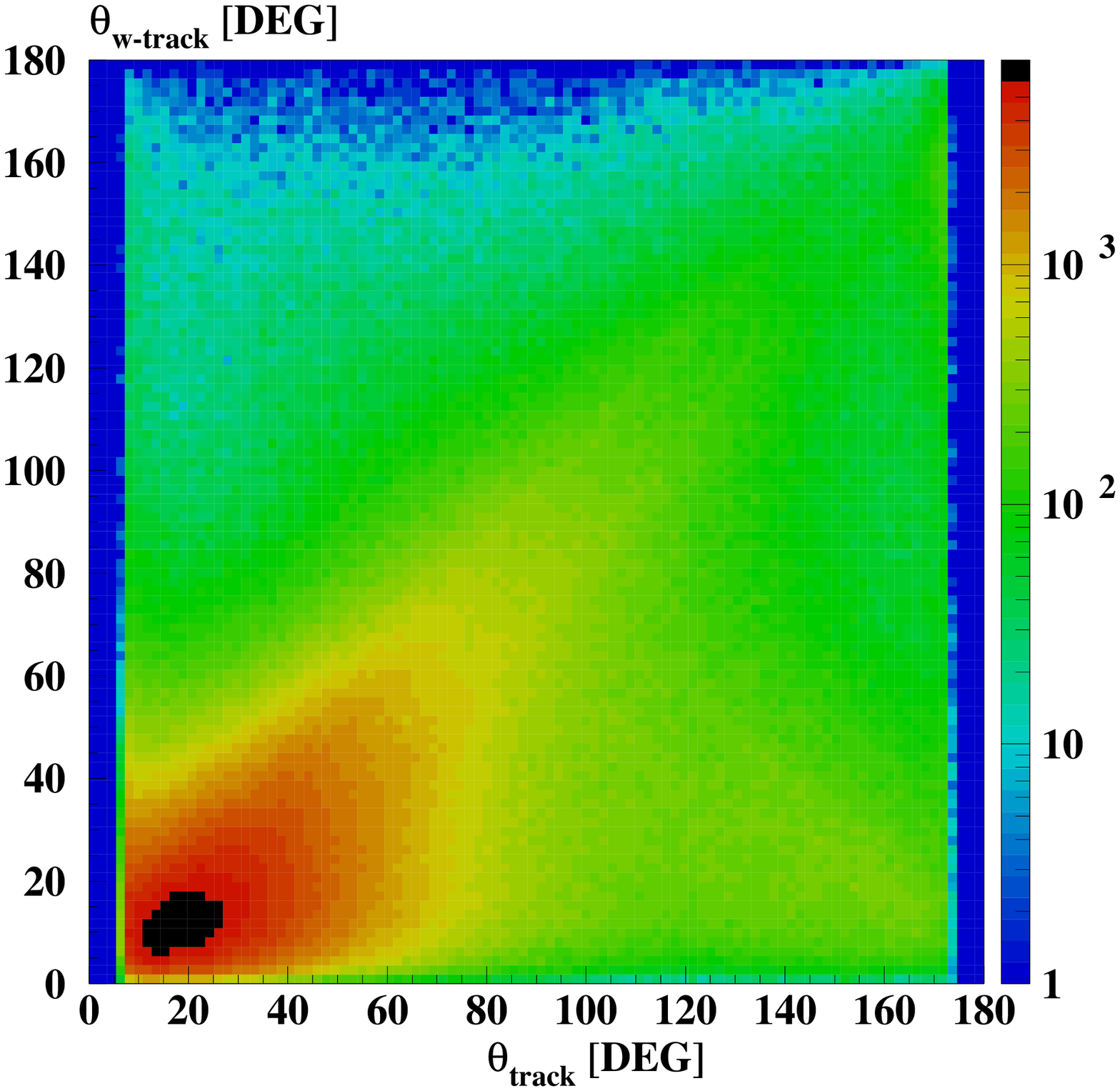}
\hfill
\includegraphics[width=0.48\linewidth,bb=0 5 567 542,clip]{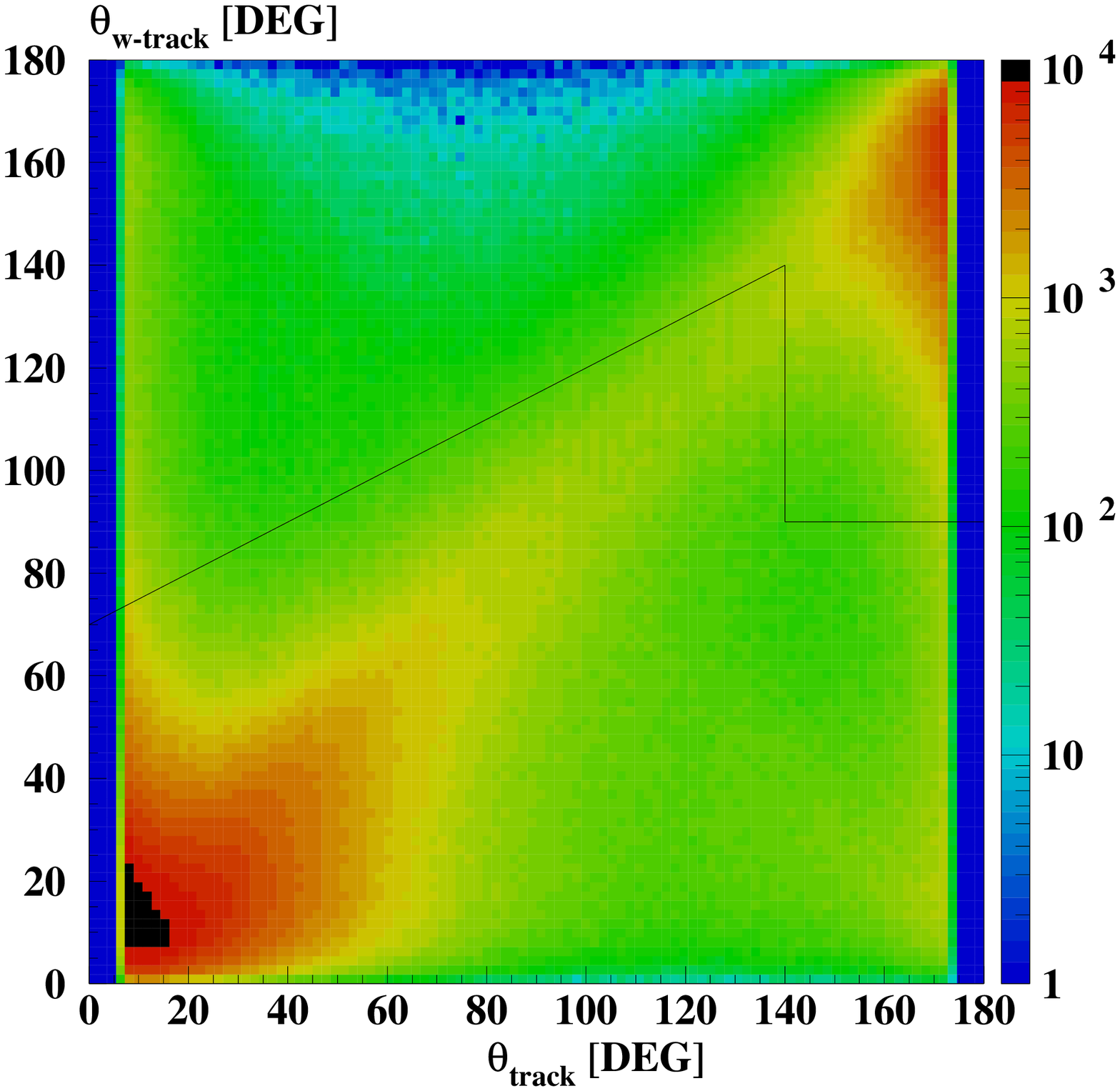}
\caption[bla]{Angle of the energy flow objects with the beam axis versus their
  angle with the reconstructed W direction for signal only (left,a) and for
  signal plus pileup (right,b). The tracks above the line shown in b are
  rejected in the analysis.}
\label{fig:f8}
\end{center}
\end{figure}

A reconstruction of the angle of each EFO with respect to the \textit{z}-axis
and the angle between the EFO and the flight direction of the reconstructed W
(Fig.~\ref{fig:f8}), makes it possible to distinguish further between signal
and pileup EFOs. EFOs are rejected if they are
positioned in the area shown in Fig.~\ref{fig:f8}\,b.

\begin{figure}[p]
\begin{center}
\epsfxsize=2.5in
\epsfysize=2.5in
\epsfbox{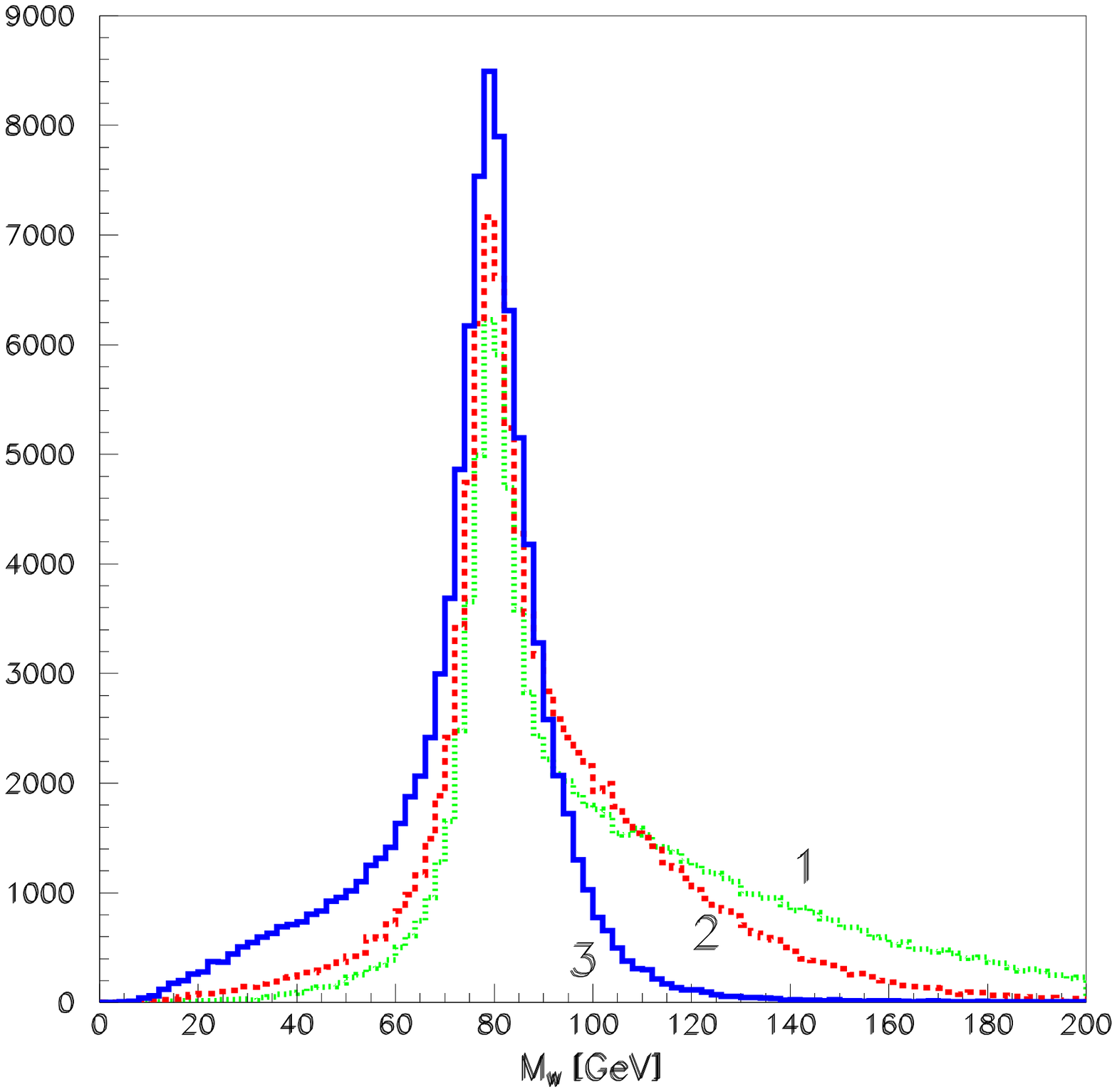}
\hspace{.25in}
\epsfxsize=2.5in
\epsfysize=2.5in
\epsfbox{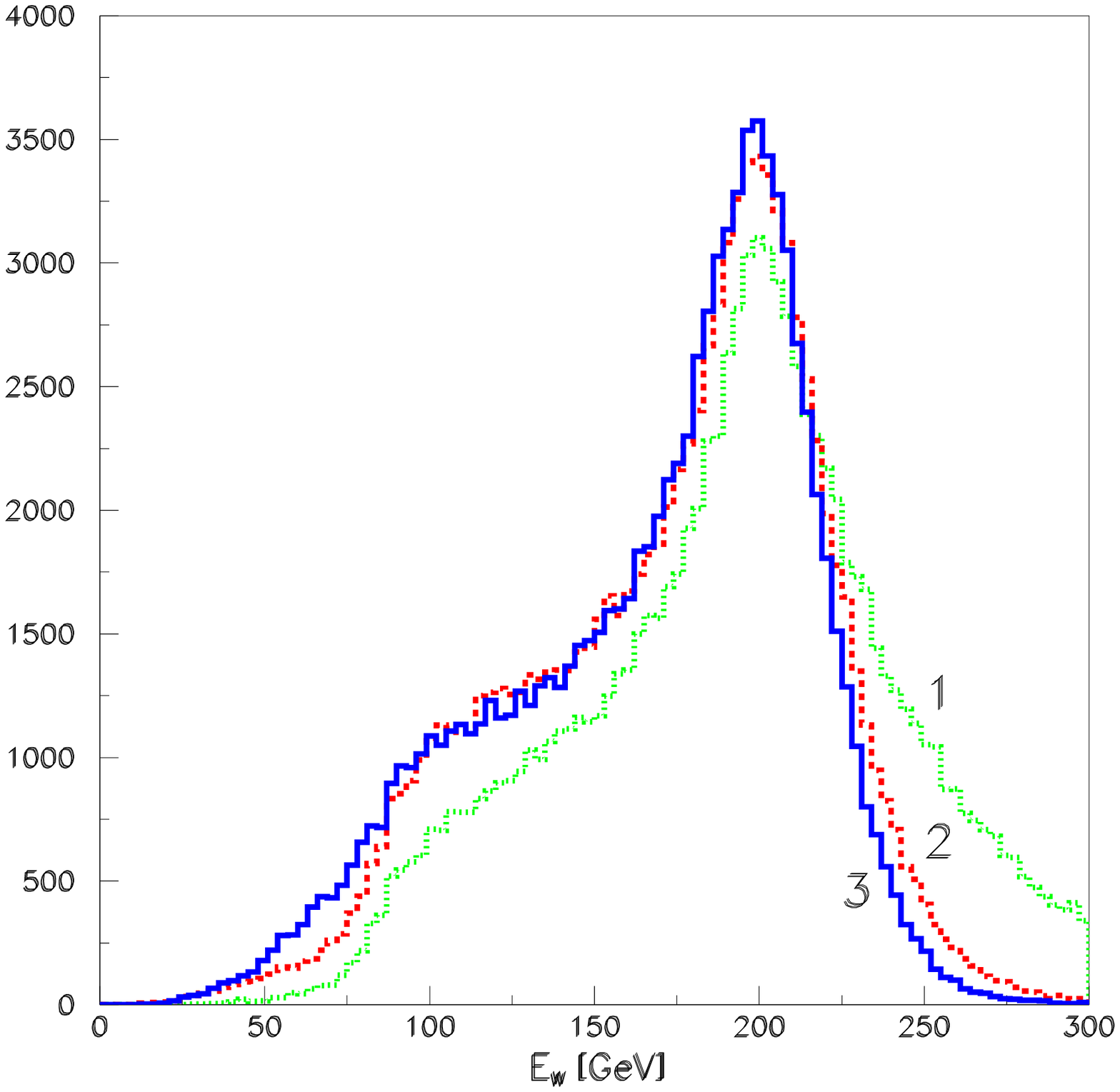}
\caption[bla]{Mass and energy distributions of the reconstructed W for the 
  real mode during the different steps in the EFO rejection; 
  initial shape (\textit{green-1}), after
  the track rejection using \textit{$I_{Z}$} (\textit{red-2}) and final shape
  (\textit{blue-3}). \textit{Left} (a): W mass distributions. \textit{Right}
  (b): W energy distributions.}
\label{fig:f9}
\end{center}
\end{figure}

\begin{figure}[p]
\begin{center}
\epsfxsize=2.5in
\epsfysize=2.5in
\epsfbox{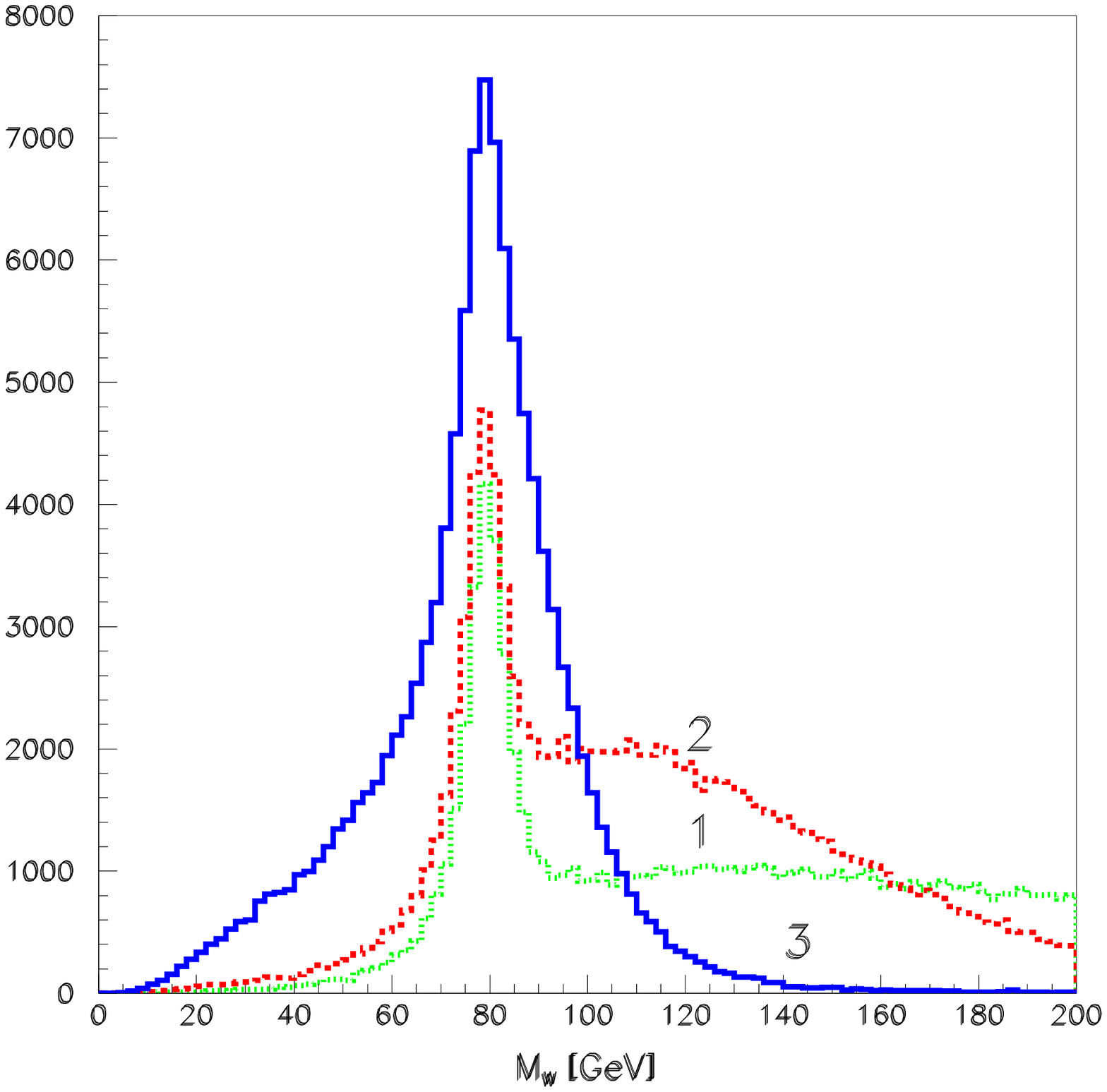}
\hspace{.25in}
\epsfxsize=2.5in
\epsfysize=2.5in
\epsfbox{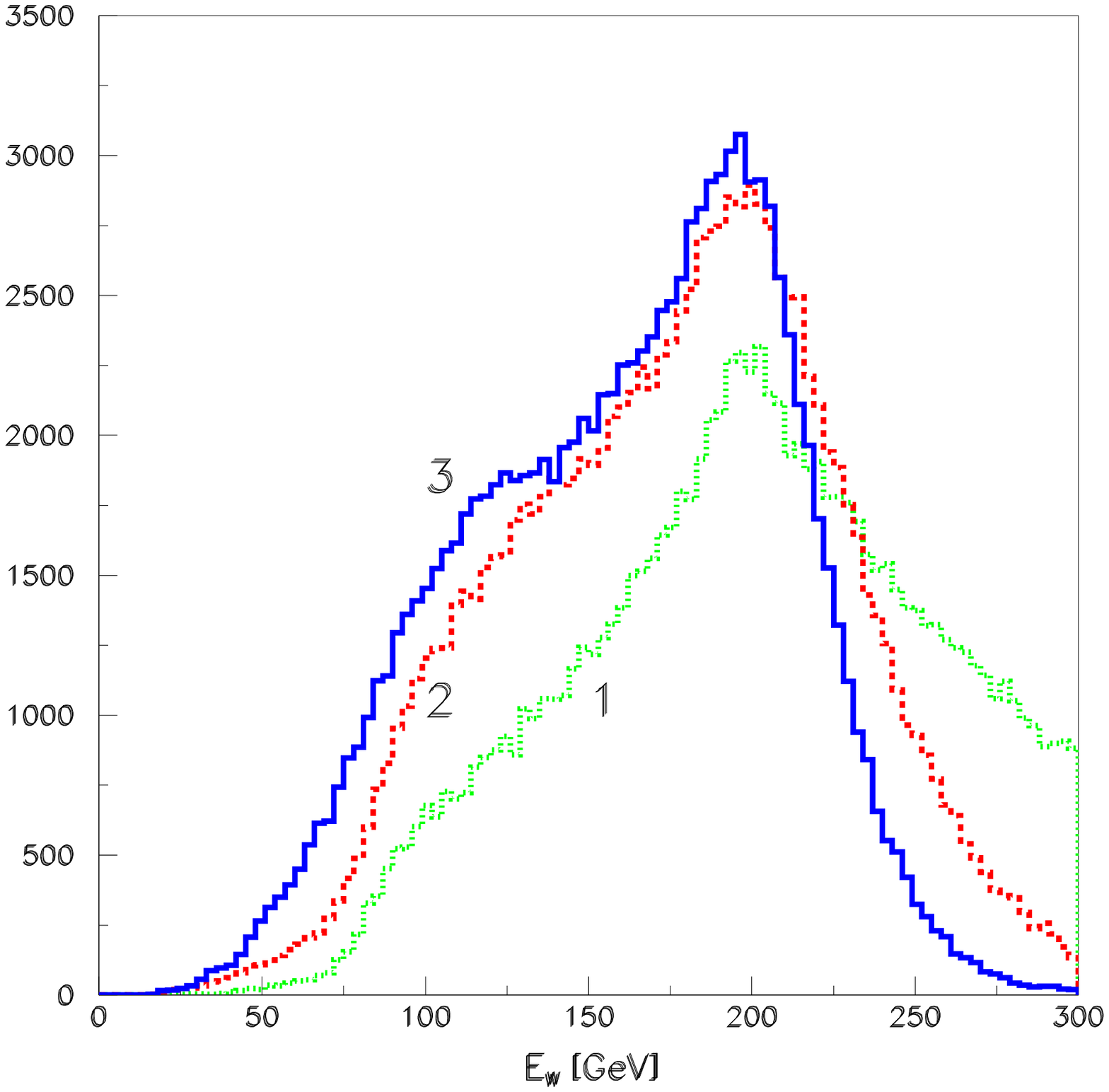}
\caption[bla]{Same a Fig.~\ref{fig:f9} for the parasitic mode.}
\label{fig:f10}
\end{center}
\end{figure}

The different steps during the separation procedure for the real and parasitic
${e}{\gamma}$-mode are shown in Fig.~\ref{fig:f9} and in Fig.~\ref{fig:f10}.
The shapes of the W distributions are restored, increasing the efficiency but
getting worse resolutions.
\par
\begin{figure}[htb]
\begin{center}
\epsfxsize=2.5in
\epsfysize=2.5in
\epsfbox{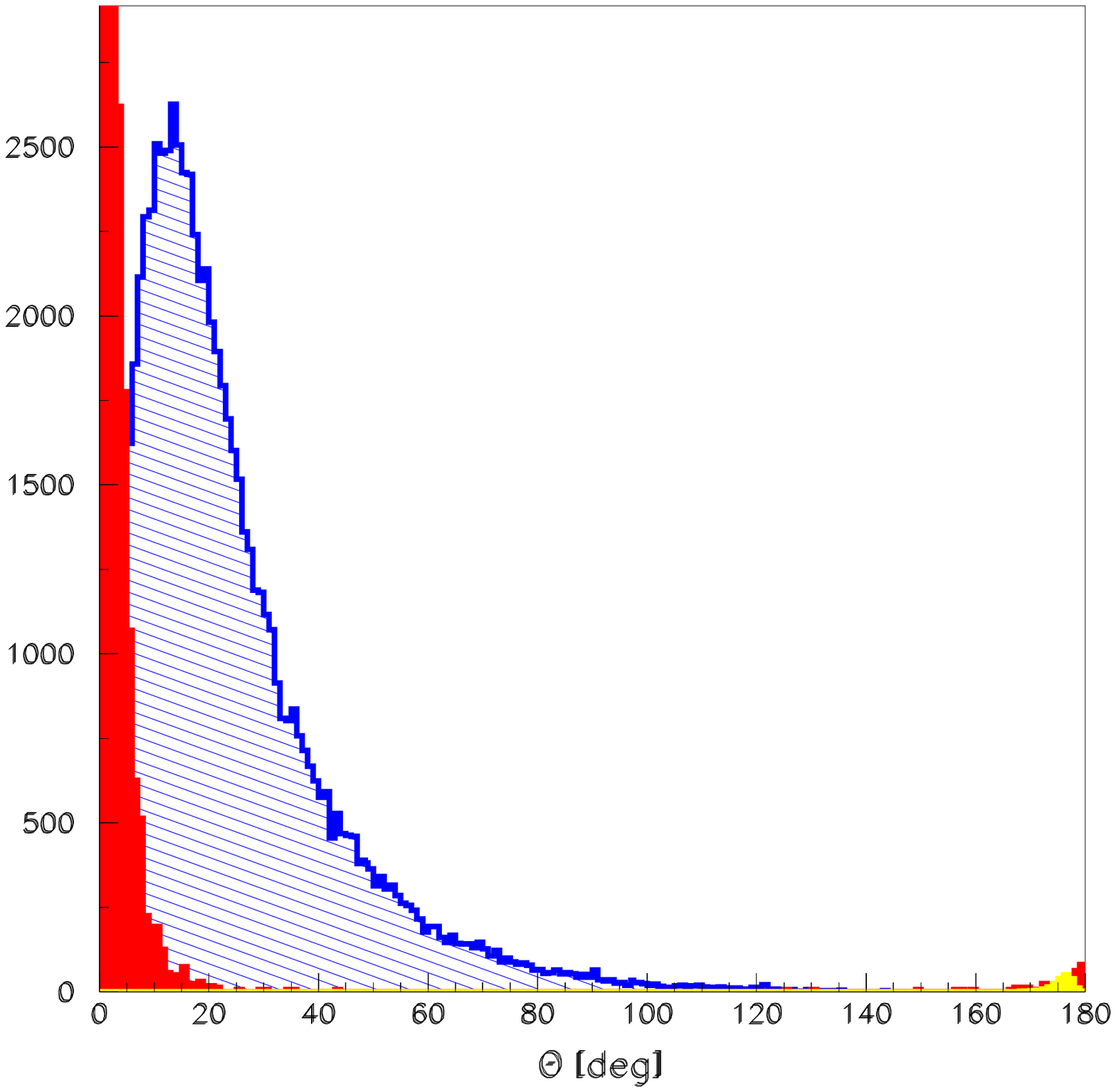}
\hspace{.25in}
\epsfxsize=2.5in
\epsfysize=2.5in
\epsfbox{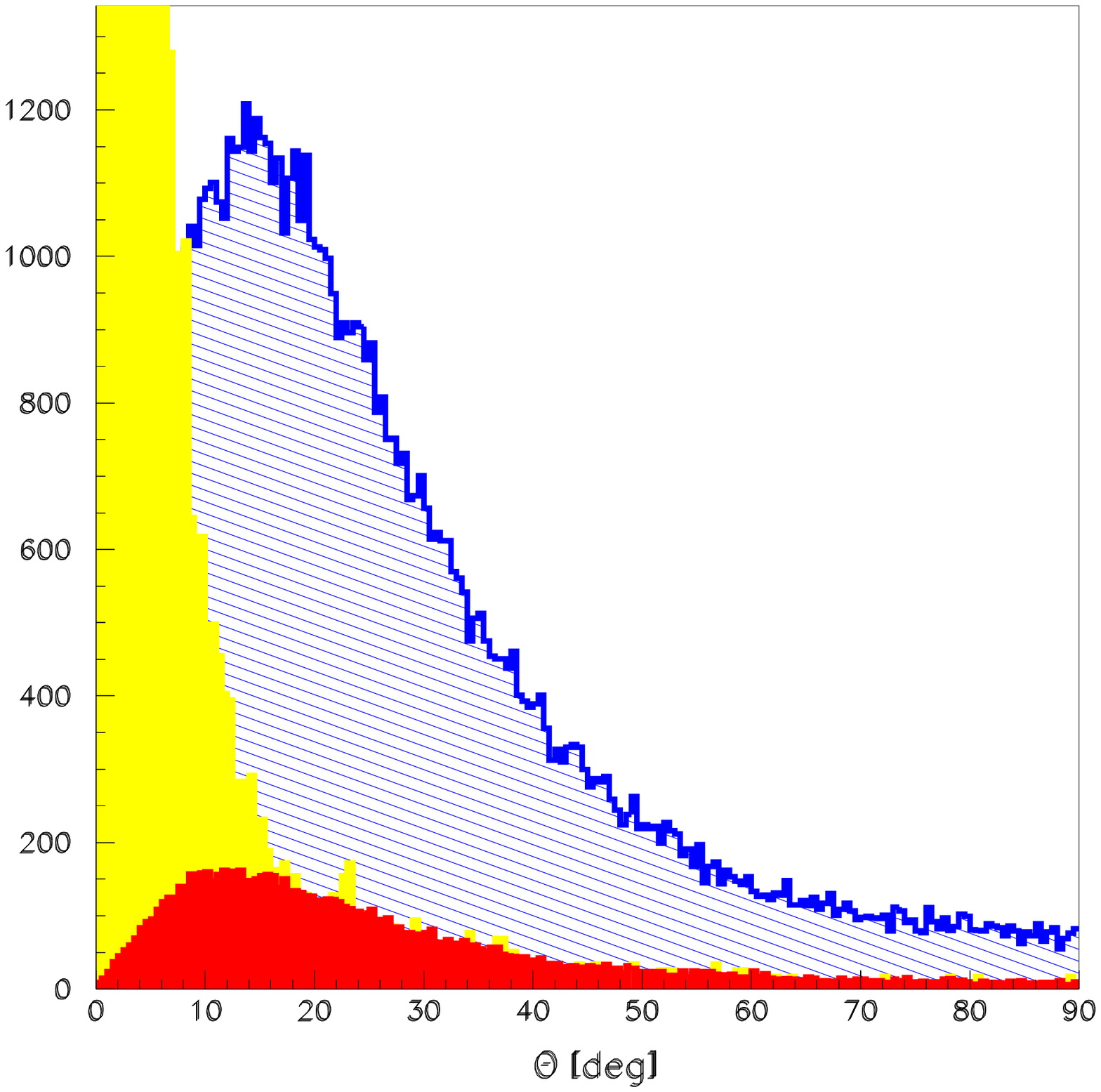}
\caption[bla]
{Signal and background distributions for $e\gamma\rightarrow W\nu$ as a
  function of the W production angle. The different processes are normalised
  to the same luminosity. The \textit{blue} (\textit{hatched}) area represents
  the signal. \textit{Left} (a): Real $e\gamma$ mode. The \textit{red}
  (\textit{dark}) contribution correspond to
  ${\gamma}(e^-){\gamma}{\rightarrow}{q}{\bar{q}}$ processes
  and the \textit{yellow} one (\textit{light}) correspond to $e \gamma
  \rightarrow eZ$. \textit{Right} (b): Parasitic $e\gamma$ mode. The
  \textit{red} (\textit{dark}) contribution correspond to
  ${\gamma}{\gamma}{\rightarrow}{W}{W}$ while the \textit{yellow} one
  (\textit{light}) correspond to
  ${\gamma}{\gamma}{\rightarrow}{q}{\bar{q}}$ processes.}
\label{fig:f11}
\end{center}
\end{figure}

In order to separate the signal events from the background the events with a
number of EFOs larger than 10 and a number of charged tracks larger than 5 are
accepted only. We also applied in addition to the vetoes on high energy and
isolated leptons cuts on two reconstructed variables, the energy ($100\,{\rm
  GeV} - 250\,{\rm GeV}$) and the mass ($60\,{\rm GeV} - 100\,{\rm GeV}$) of
the reconstructed W boson. The final angular distributions of signal and
background events for both ${e}{\gamma}$-modes are shown in Fig.~\ref{fig:f11}.

The efficiency obtained for the real mode is $73\,{\%}$ with a purity of
$64\,{\%}$. In the parasitic mode, due to the fact that the pileup is
larger than in the case of the real mode, the efficiency is $66\,{\%}$
with a purity is $49\,{\%}$. Background events are mostly distributed
close to the beam pipe and an additional cut on the W production angle is
applied in order to increase the purity of the signal in both modes. Events in
the region below $5^{\circ}$ are rejected leading to a purity of 
$95\,{\%}$ for the real mode and $72\,{\%}$ for the parasitic mode. This
cut has only a small influence on the signal resulting in efficiencies of
$70\,{\%}$ and $63\,{\%}$ for real and parasitic mode, respectively.
%%%%%%%%%%%%%%%%%%%%%%%%%%%%%%%%%%%%%%%%%%%%%%%%%%%%%%%%%%%%%%%%%%%
\section{Fit Method and Error Estimations}
\begin{figure}[htb]
\begin{center}
\epsfxsize=2.5in
\epsfysize=2.5in
\epsfbox{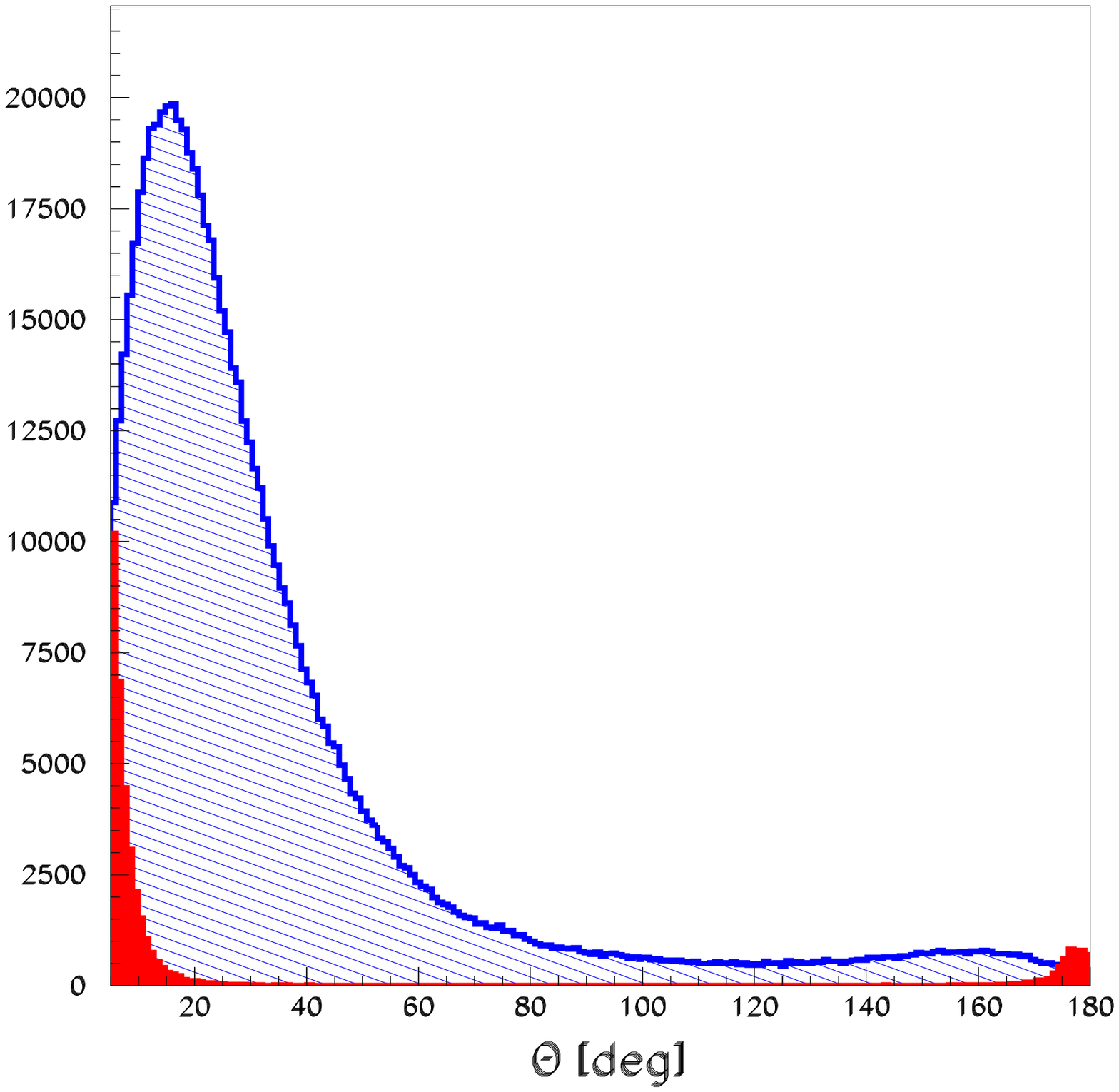}
\hspace{.25in}
\epsfxsize=2.5in
\epsfysize=2.5in
\epsfbox{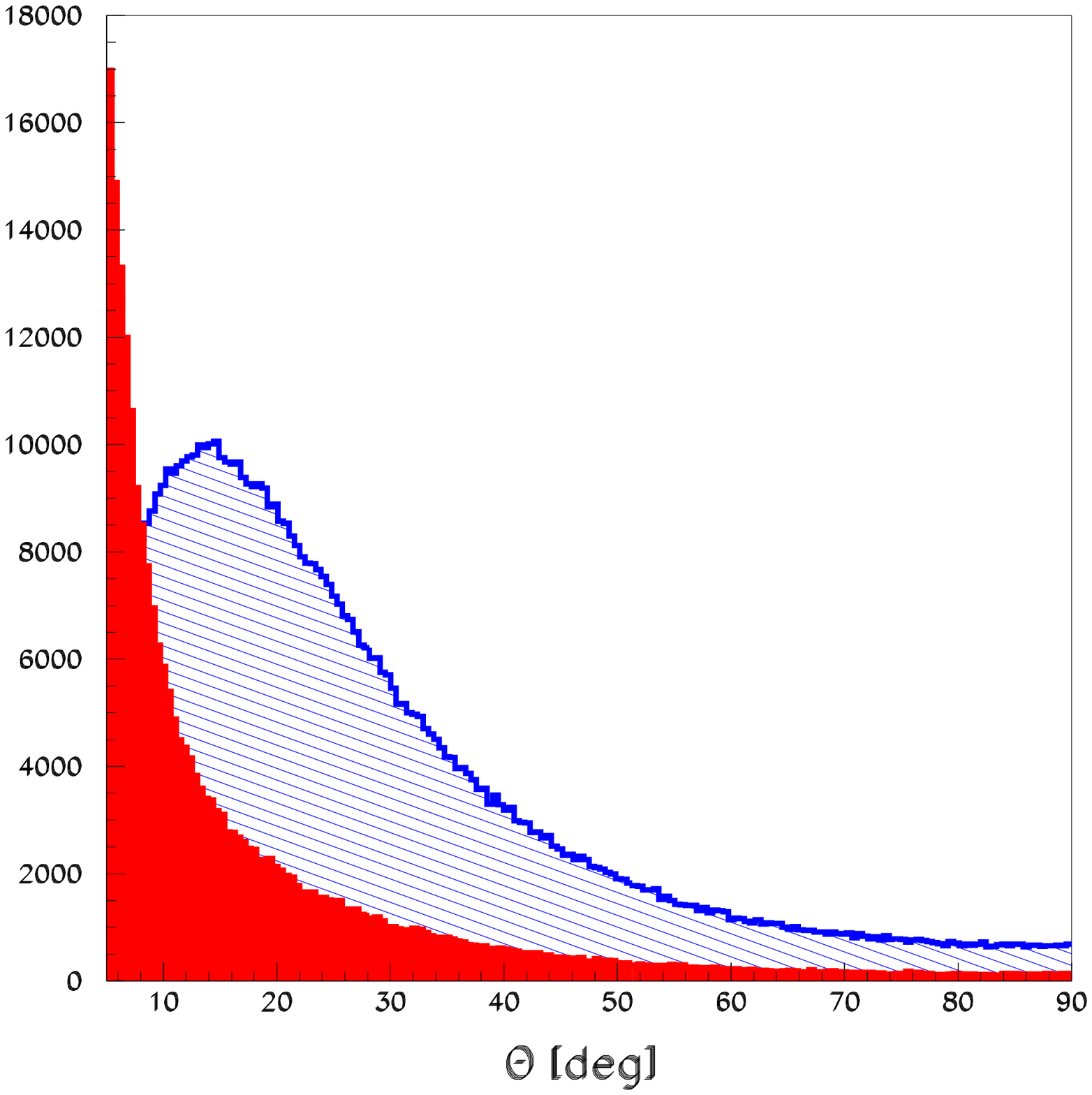}
\caption[bla]
{Polar angle distributions of reconstructed Ws for the real (\textit{left}) and
  parasitic (\textit{right}) mode. The signal events are coloured with
  \textit{blue} (\textit{hatched}) while the \textit{red} (\textit{full})
  distributions correspond to the background events.}
\label{fig:f12}
\end{center}
\end{figure}

%%%%%%%%%%%%%%%%%%%%%%%%%%%%%%%%%%%%%%%%%%%%%%%%%%%%%%%%%%%%%%%%%%%
For the extraction of the triple gauge couplings from the reconstructed
kinematical variables a ${\chi^{2}}$ fit is used. A sample of ${10^{6}}$ SM
signal events is generated with WHIZARD and passed trough the detector
simulation. The number of events obtained after the detector and after all
cuts (Fig.~\ref{fig:f12}) is normalised to the number of events we expect
after one year of running of an $e \gamma$-collider.
\par
Each event is described reconstructing three kinematical variables - the W
production angle with respect to the ${e^{-}}$ beam direction, the W polar
decay angle $\cos{\theta_{1}}$ (angle of the fermion with respect to the W
flight direction measured in the W rest frame) and the azimuthal decay angle
$\phi$ of the fermion with respect to a plane defined by W and the beam axis.
The polar decay angle, $\cos{\theta_{1}}$ is sensitive to the different W
helicity states and the azimuthal angle, $\phi$ to the interference between
them. In hadronic W-decays the up- and down-type quarks cannot be separated so
that only $| \cos \theta_1 |$ is measured.  The matrix element calculations
from WHIZARD are used to obtain weights to reweight the angular distributions
as functions of the anomalous TGCs. Each Monte Carlo SM event is weighted by a
weight:
%\begin{equation}
\[
{R({\Delta}{\kappa}_{\gamma},{\lambda}_{\gamma}) =
1+A{\Delta}{\kappa}_{\gamma}+B{{{\Delta}{\kappa}_{\gamma}}^{2}}+
C{\lambda}_{\gamma}+D{{{\lambda}_{\gamma}}^{2}}+
E{\Delta}{\kappa}_{\gamma}{\lambda}_{\gamma}}
\]
%\label{eq:r}
%\end{equation}
where ${\Delta}{\kappa}_{\gamma}$ and ${\lambda}_{\gamma}$ are the free
parameters. The function ${R({\Delta}{\kappa}_{\gamma},{\lambda}_{\gamma})}$
describes the quadratic dependence of the differential cross-section on the
coupling parameters and it is obtained in the following way: using SM events
(${\Delta}{\kappa}_{\gamma}={\lambda}_{\gamma}=0$) we recalculated the matrix
elements of the events for a set of five different combinations of
${\Delta}{\kappa}_{\gamma}$ and ${\lambda}_{\gamma}$ values (Table
\ref{tab:1}).
\begin{table}[h]
\begin{center}
\begin{tabular}{|c|c|c|c|c|c|} \hline
& $R_1$& $R_2$& $R_3$& $R_4$& $R_5$ \\ \hline
${\Delta}{\kappa}_{\gamma}$ & 0 & 0 & +0.001 & -0.001  & +0.001 \\ \hline
${\lambda}_{\gamma}$ & +0.001 & -0.001& 0 & 0 & +0.001 \\ \hline
\end{tabular}
\end{center}
\caption{${\Delta}{\kappa}_{\gamma}$, ${\lambda}_{\gamma}$ 
values used to calculate the reweighting coefficients.}
\label{tab:1}
\end{table}
\par
The resulting recalculated events carry a weight which is given by the ratio
of the new matrix element values compared to the SM ones ($R_{i}$). The
particle momenta are left unchanged. According to the chosen
${\Delta}{\kappa}_{\gamma},{\lambda}_{\gamma}$ combinations from Table
\ref{tab:1} one gets:
\begin{eqnarray*}
 R_{1} & = & 
 1+C\mid{\lambda}_{\gamma}\mid+D\mid{\lambda}_{\gamma}^{2}\mid,\\
 R_{2} & = & 
 1-C\mid{\lambda}_{\gamma}\mid+D\mid{\lambda}_{\gamma}^{2}\mid,\\
 R_{3} & = & 
 1+A\mid{\Delta}{\kappa}_{\gamma}\mid+B\mid{\Delta}{\kappa}_{\gamma}^{2}\mid,\\
 R_{4} & = & 
 1-A\mid{\Delta}{\kappa}_{\gamma}\mid+B\mid{{{\Delta}
 {\kappa}_{\gamma}}^{2}\mid},\\
 R_{5} & = & 
 1+A\mid{\Delta}{\kappa}_{\gamma}\mid+B\mid{{{\Delta}
 {\kappa}_{\gamma}}^{2}\mid}+C\mid{\lambda}_{\gamma}\mid+
  D\mid{{{\lambda}_{\gamma}}^{2}\mid}+E\mid{\Delta}
  {\kappa}_{\gamma}\mid\mid{\lambda}_{\gamma}\mid, 
\end{eqnarray*}
where $|{\Delta}{\kappa}_{\gamma}|$=$|{\lambda}_{\gamma}|$=0.001. The
coefficients \textit{A,B,C,D,E} are deduced for each event from the previous
five equations.
%\begin{eqnarray*}
% D & = & \frac{(M_{1}+M_{2})}{0.002},\\
% C & = & M_{1}-0.001\cdot{D},\\
% B & = & \frac{(M_{3}+M_{4})}{0.002},\\
% A & = & M_{3}-0.001\cdot{B},\\
%% E & = & \frac{(M_{5}-A-C)}{0.001}-(B+D),
%\end{eqnarray*}
%with $M_{i} = \frac{R_{i}-1}{0.001}$.
Four-dimensional ($\cos{\theta}$,$\cos{\theta_{1}},\phi$, energy) event
distributions are fitted with MINUIT \cite{min},
minimizing the ${\chi^{2}}$ as a function of $\kappa_\gamma$ and
$\lambda_\gamma$ taking the SM Monte Carlo sample as ``data'':
\[
\chi^{2} = \sum_{i,j,k,l}
\frac{ \left( z\cdot N^{SM}(i,j,k,l)- n \cdot z \cdot 
N^{{\kappa}_{\gamma},{\lambda}_{\gamma}}(i,j,k,l) \right)^{2}} 
{z \cdot \sigma^{2}(i,j,k,l)} +\frac{(n-1)^{2}}{(\Delta L^{2})}
\]
where \textit{i, j} and \textit{k} run over the reconstructed angular
distributions, \textit{l} runs over the reconstructed W boson energy,
$N^{SM}(i,j,k,l)$ are the ``data'' which correspond to the SM Monte Carlo
sample, $N^{{\kappa}_{\gamma},{\lambda}_{\gamma}}(i,j,k,l)$ is the event
distribution weighted by the function
$R({\Delta}{\kappa}_{\gamma},{\lambda}_{\gamma})$ and
$\sigma(i,j,k,l)=\sqrt{N^{SM}(i,j,k,l)}$.  The factor ${z}$ sets the number of
signal events to the expected one after one year of running of an $e
\gamma$-collider. In case where the background is included in the fit ${z}$
defines the sum of signal and background events and ${n}\cdot
N^{{\kappa}_{\gamma},{\lambda}_{\gamma}} \rightarrow [{n} \cdot
N_{signal}^{{\kappa}_{\gamma},{\lambda}_{\gamma}}+N_{bck}]$. The number of
background events is normalised to the effective W production cross-section in
order to obtain the corresponding number of background events after one year
of running of an $e \gamma$-collider. 
It is assumed that the total normalisation
(efficiency, luminosity, electron polarisation) is only known with a relative
uncertainty $\Delta L$. To do this $n$ is taken as a free parameter in the fit
and constrained to unity with the assumed normalisation uncertainty. Per
construction the fit is bias-free and thus returns always exactly the SM as
central values.
%%%%%%%%%%%%%%%%%%%%%%%%%%%%%%%%%%%%%%%%%%%%%%%%%%%%%%%%%%%%%%%
\par
Table \ref{tab:t2} shows the estimated statistical errors we expect for the
different couplings at ${\sqrt{s_{ee}}=500}\,{\rm GeV}$ for
two-parameter\footnote{A two-parameter fit means that both couplings are
  allowed to vary freely as well as the normalisation \textit{n}.}
four-dimensional (4D) fit at detector level, with and
without pileup. In this estimation the cut of $5^{\circ}$ is not applied.
Table \ref{tab:t3} contains the statistical errors obtained
together with background events applying the cut on the W production angle of
$5^{\circ}$.
\begin{table}[h]
\begin{center} 
\begin{tabular}{|l|c|c|c||c|c|c|} \hline
 & \multicolumn{3}{|c||}{without pileup} & \multicolumn{3}{c|}{with pileup}\\
\hline
${{\Delta}L}$ & 1$\%$ & 0.1$\%$ & 0 &
 1$\%$ & 0.1$\%$ & 0 \\ \hline\hline
${\Delta}{\kappa}_{\gamma}{\cdot}10^{-3}$ & 3.4/4.0 & 1.0/1.0 & 0.5/0.5 & 
 3.5/4.5 & 1.0/1.0 & 0.5/0.5 \\ 
\hline
${\Delta}{\lambda}_{\gamma}{\cdot}10^{-4}$ & 4.9/5.5 & 4.5/5.2 & 4.5/5.1 & 
5.2/6.7 & 4.9/6.4 & 4.9/6.4 \\ \hline
\end{tabular}
\end{center}
\caption{
Estimated statistical errors for ${\kappa}_{\gamma}$ and 
${\lambda}_{\gamma}$ from the two-parameter 4D fit at detector level
for the real/parasitic $e\gamma$ mode without and 
with pileup.}
\label{tab:t2}
\end{table}
\begin{table}[htb]
\begin{center}
\begin{tabular}{|l|c|c|c|} \hline
 & \multicolumn{3}{|c|}{pileup+background} \\ \hline
${{\Delta}L}$ & 1$\%$ & 0.1$\%$ & 0 \\ \hline\hline
${\Delta}{\kappa}_{\gamma}{\cdot}10^{-3}$ & 3.6/4.8 & 1.0/1.1 & 0.5/0.6 \\ 
\hline
${\Delta}{\lambda}_{\gamma}{\cdot}10^{-4}$ & 5.2/7.0 & 4.9/6.7 & 4.9/6.7 \\ 
\hline
\end{tabular}
\end{center}
\caption{Estimated statistical errors for ${\kappa}_{\gamma}$ and 
${\lambda}_{\gamma}$ from the two-parameter 4D fit at detector level 
for the real/parasitic $e$$\gamma$ mode with pileup and background events.}
\label{tab:t3}
\end{table}
\par
The main error on ${\kappa}_{\gamma}$ comes from the luminosity measurement
while ${\lambda}_{\gamma}$ is not sensitive to that uncertainty. The two
different $e \gamma$ modes give the same estimation for
${\Delta}{\kappa}_{\gamma}$ while ${\Delta}{\lambda}_{\gamma}$ is more
sensitive to the different modes. The difference in the estimated
${\Delta}{\lambda}_{\gamma}$ for two modes is a consequence of the ambiguity
in the W production angle which is present in the parasitic mode\footnote{
  In a parasitic mode only $|\cos\theta|$ can be reconstructed.} 
and due to the fact
that the distance between the conversion region and the interaction point
is larger in the real mode than in the parasitic mode. 
A smaller distance between the conversion and the interaction region
increases the luminosity at the price of a broader energy spectrum.
That decreases the sensitivity of the ${\lambda}_{\gamma}$ measurement.
\par
\begin{figure}[h]
\begin{center}
\epsfxsize=3.0in
\epsfysize=3.0in
\epsfbox{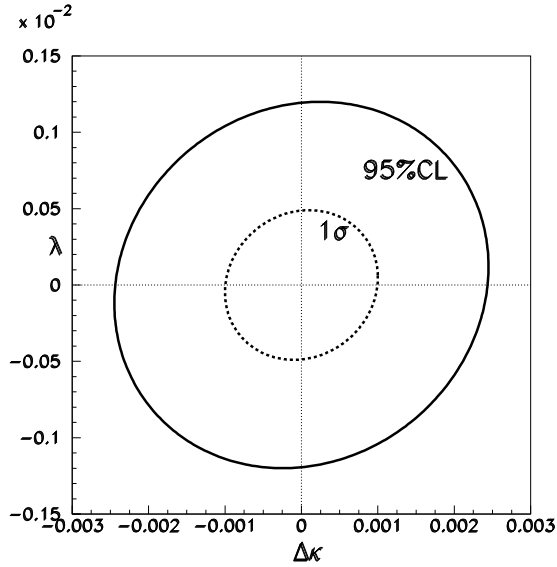}
\caption[bla]
{95$\%$ CL and 1$\sigma$ contours in the
  ${\Delta}{\kappa}_{\gamma}-{\lambda}_{\gamma}$ plane obtained from the 4D
  fit for ${{\Delta}L}=0.1\%$.}
\label{fig:f13}
\end{center}
\end{figure}

The pileup contribution is larger in the parasitic than in the real mode
and therefore it influences the W distributions (energy and angular)
more than in the real mode. This leads to a decrease in sensitivity for
${\lambda}_{\gamma}$ of $\sim 10\,{\%}$ in the real and of $\sim
25\,{\%}$ in the parasitic mode\footnote{
  All comparisons are done assuming ${{\Delta}L}=0.1\%$.} 
while the influence on
${\Delta}{\kappa}_{\gamma}$ is negligible.  The influence of the
background is not so stressed as it is for the pileup. In the real
mode it is almost negligible while it contributes to the parasitic
mode decreasing the sensitivity of ${\lambda}_{\gamma}$ by less than
$5\,{\%}$.  The contour plot in ${\Delta}{\kappa}_{\gamma} -
{\lambda}_{\gamma}$ plane, shown in Fig.~\ref{fig:f13} is based on the
results given in Table \ref{tab:t3} assuming a normalisation error
of 0.1${\%}$.
\par
The correlation between the fit parameters ${\Delta}{\kappa}_{\gamma}$ and
${\Delta}{\lambda}_{\gamma}$ is found to be negligible and it is shown in
Table \ref{tab:t4} while ${\Delta}{\kappa}_{\gamma}$ strongly depends on
\textit{n}.
\begin{table}[htb]
\begin{center}
\begin{tabular}{|l|c|c|c|} \hline
& \multicolumn{3}{c|}{pileup+background}\\ \hline
& ${\Delta}{\kappa}_{\gamma}$ & $\textit{n}$ & ${\Delta}{\lambda}_{\gamma}$ \\ \hline

${\Delta}{\kappa}_{\gamma}$ & 1.000 & -0.857 & 0.122 \\ \hline
$\textit{n}$ & -0.857 & 1.000 & -0.094 \\ \hline
${\Delta}{\lambda}_{\gamma}$ & 0.122 & -0.094 & 1.000 \\ \hline
\end{tabular}
\end{center}
\caption{Correlation matrix for the two-parameter fit 
(${{\Delta}L}=0.1\%$).}
\label{tab:t4}
\end{table}
%%%%%%%%%%%%%%%%%%%%%%%%%%%%%%
\subsection{Systematic Errors}
%%%%%%%%%%%%%%%%%%%%%%%%%%%%%%
Due to the large W production cross-sections and achievable luminosities at
the PC the statistical errors are comparable with those estimated for the
$e^{+}e^{-}$-collider \cite{menges} and the main source of error comes from
the systematics. Some sources of systematic errors have been investigated,
assuming ${{\Delta}L}=0.1\%$. It was found that the largest uncertainty in
${\kappa}_{\gamma}$ comes from uncertainties on the photon beam polarisations.
Contrary to the $e^+ e^-$ case the luminosity and polarisation measurements
are not independent. The dominant polarisation state ($J_z = 3/2$) can be
measured accurately with $e \gamma \rightarrow e \gamma$ while the suppressed
one ($J_z = 1/2$) can only be measured with worse precision e.g. using 
$e Z \rightarrow e Z$ \cite{polar}.
To estimate the uncertainty on the TGCs therefore the dominant $J_z = 3/2$
part is kept constant while the $J_z = 1/2$ part is changed by 10\%,
corresponding to a 1\% polarisation uncertainty for ${\cal P}_\gamma=0.9$.
This leads to a polarisation uncertainty of $0.005$ for $\kappa_\gamma$,
corresponding to five times the statistical error while the uncertainty on
$\lambda_\gamma$ is negligible.
The photon polarisation thus
needs to be known to 0.1$\%$ - 0.2$\%$ so that ${\kappa}_{\gamma}$ is not
dominated by this systematic error.
\par
In order to estimate the error coming from the W mass measurement we
recalculated the data sample with ${M_{W}}$ decreased/increased by $50\,{\rm
  MeV}$ (the expected ${\Delta}{M_{W}}$ at LHC is $\sim 15$ MeV) reweighting
the SM events. The nominal W mass used for Monte Carlo sample was
$M_{W}=80.419\,{\rm GeV}$. As a result of the recalculation we get the ratios
of matrix element values corresponding to the nominal W mass and the mass
$M_{W}^{'}= M_W \pm \Delta M_{W}$. The Monte Carlo sample (MC) is weighted by
this ratio and fitted as fake data leaving the reference distributions
%\footnote{MC sample with a nominal mass.} 
unchanged. The resulting shift for TGCs is
of the order of the statistical error for both coupling parameters
for $\Delta M_W = 50\,{\rm MeV}$ and thus negligible with an improved W-mass
measurement. 
\par
At the PC the field of the laser wave at the conversion region is very
strong and the high energy electron or photon can interact
simultaneously with several laser photons. These are nonlinear QED
effects that influence the Compton spectra of the scattered photons in
such a way that increasing the nonlinearity $\xi^{2}$ \cite{tdr6} the
Compton spectrum becomes broader and shifted to lower energies.  To
estimate the error that comes from this effect the laser power is
decreased changing $\xi^{2}$ from 0.3 $\rightarrow$ 0.15, increasing
the peak energy by 2.5\%.  The ratio of the two Compton spectra is
used as a weight function to obtain the ``data'' sample from the MC
events.  The sample data obtained in that way are fitted leaving the
reference distributions unchanged. It was found that the beam energy
uncertainty influences the measurement of the coupling parameters only
via the normalisation $n$, and the errors ${\Delta}{\kappa}_{\gamma}$
and ${\Delta}{\lambda}_{\gamma}$ are considered as negligible since
the value of $n$ is accessible from the luminosity measurement.
\par
The estimated systematic error for $\kappa_\gamma$ from background
uncertainties is smaller than the statistical error if the background cross
section is know to better than 3\% in the real mode and 1\% in the parasitic
mode. For $\lambda_\gamma$ the background needs to be known only to 10\% in the
parasitic mode while there are practically no restrictions in the real mode.
%%%%%%%%%%%%%%%%%%%%%%%%
\section{Conclusions}
%%%%%%%%%%%%%%%%%%%%%%%%
A future high energy $e \gamma$ collider provides an excellent opportunity to
measure the gauge couplings between a W-pair and a photon. These couplings can
be obtained without ambiguities from quartic or ZWW couplings.  The expected
precision is $10^{-3}$ for $\kappa_\gamma$ and $10^{-4}$ for $\lambda_\gamma$.
While $\kappa_\gamma$ can be measured somewhat better in $e^+ e^-$
\cite{menges} the $e \gamma$ collider seems to be the best place for a
$\lambda_\gamma$ measurement.
\section*{Acknowledgements}
%%%%%%%%%%%%%%%%%%%%%%%%

We would like to thank Wolfgang Kilian for many useful advises in 
the usage of WHIZARD.
%%%%%%%%%%%%%%%%%%%%%%%%

%%%%%%%%
%\newpage
\boldmath

\end{document}